\documentclass[usenatbib]{mn2e}
\usepackage{epsfig}
\usepackage{amsmath,amssymb}
\usepackage{color}
\usepackage{natbib}

  \newcommand{\Msolar}{\mbox{\,$\rm M_{\odot}$}}        
  \newcommand{\iso}[2]{\mbox{$^{#1}{\rm #2}$}}           
  \def\simge{\mathrel{\raise1.16pt\hbox{$>$}\kern-7.0pt
    \lower3.06pt\hbox{{$\scriptstyle \sim$}}}}           
  \def\simle{\mathrel{\raise1.16pt\hbox{$<$}\kern-7.0pt
    \lower3.06pt\hbox{{$\scriptstyle \sim$}}}}           
\begin{document}

    \title[Surface abundances in EHe and R\,CrB stars]
         {Double white dwarf mergers and elemental surface abundances in 
          extreme helium and R\,Coronae Borealis stars}

\author[C.S.Jeffery, A.I.Karakas \& H.Saio]
       {C.S.Jeffery$^{1,2}$\thanks{E-mail: csj@arm.ac.uk}, 
A.I.Karakas$^3$\thanks{E-mail: akarakas@mso.anu.edu.au}
\& H.Saio$^4$\thanks{E-mail: saio@astr.tohoku.ac.jp}  
\\
$^1$Armagh Observatory, College Hill, Armagh BT61 9DG, Northern Ireland\\
$^2$School of Physics, Trinity College Dublin, Dublin 2, Ireland\\
$^3$Research School of Astronomy \& Astrophysics, Mt Stromlo Observatory, Canberra, Australia \\
$^4$Astronomical Institute, School of Science, Tohoku University, Sendai 980-8578, Japan\\
}

\date{Accepted .....
      Received ..... ;
      in original form .....}

\pagerange{\pageref{firstpage}--\pageref{lastpage}}
\pubyear{2010}
\label{firstpage}
\maketitle

%
%
\begin{abstract}
The surface abundances of extreme helium (EHe) and R Coronae Borealis
(RCB) stars are discussed in terms of a model for their origin in the merger of a
carbon-oxygen white dwarf with a helium white dwarf. The model is
expressed as a linear mixture of the individual layers of both
constituent white dwarfs, taking account of the specific evolution of
each star. In developing this recipe from previous versions,
particular attention has been given to the inter-shell abundances of
the asymptotic giant branch star which evolved to become the
carbon-oxygen white dwarf. Thus the surface composition of the merged
star is estimated as a function of the initial mass and metallicity of 
its progenitor. The question of whether additional nucleosynthesis 
occurs during the white dwarf merger has been examined by including
the results of recent hydrodynamical merger calculations which 
incorporate the major nuclear networks. 

The high observed abundances of carbon and oxygen must either
originate by dredge-up from the core of the carbon-oxygen white dwarf
during a {\it cold} merger or be generated directly by
$\alpha$-burning during a {\it hot}
merger. The presence of large quantities of $^{18}{\rm O}$ may be
consistent with both scenarios, since a significant  $^{18}{\rm O}$
pocket develops at the carbon/helium boundary in a number of our 
post-AGB models. 

The production of fluorine, neon and phosphorus in the AGB intershell
propagates through to an overabundance at the surface of the merged
stars, but generally not in sufficient quantity to match the observed 
abundances. However, the evidence for an AGB origin for these
elements, together with near-normal abundances of magnesium, 
points to progenitor stars with initial masses in the range 1.9 -- 3 \Msolar.

There is not yet sufficient understanding of the chemical
structure of CO white dwarfs, or of nucleosynthesis during a double white
dwarf merger, to discriminate the origin (fossil or prompt) of all the
abundance anomalies observed in EHe and RCB stars. 
Further work is required to quantify the expected yields of argon and
s-process elements in the AGB intershell, and to improve the predicted
yields of all elements from a {\it hot} merger.

\end{abstract}


%

\section{Introduction}
\label{intro}

Extreme helium stars (EHe: spectral types O--A), 
R\,Coronae Borealis stars (RCB: spectral types F--G) and hydrogen-deficient
carbon stars (HdC) are early- to 
late-type supergiants with atmospheres almost void of hydrogen, but
highly enriched in carbon \citep{jeffery08c,clayton96,asplund00}. 
They display an extraordinary mixture of atomic species in ratios
very different to those likely to have been established when the star
was formed. In addition to hydrogen and helium, anomalies include  
 large enrichments of nitrogen, oxygen, fluorine, 
neon and phosphorus, detections of lithium and s-process 
elements, overabundance in silicon and sulphur, together with a large
range in iron abundance \citep{jeffery96a,rao96,pandey06}. It is
commonly accepted that the hydrogen is a remnant of the outer layers
of the original star, the enriched nitrogen comes from CNO-processed
layers, helium from CNO-processed material and partially 
triple-$\alpha$-processed material, and the carbon from
3$\alpha$ reactions \citep{heber83}. There is increasing consensus
concerning the origin of other elements, where the
signature of material from the intershell (helium-rich) layers of
asymptotic giant branch stars seems unmistakable \citep{pandey06}.   

In order to interpret these abundances, it is necessary to infer some
cataclysm in the history of the star by which the hydrogen-rich
surface has been either ejected or consumed, whilst at the same time 
revealing a mixture of both CNO-processed helium, 3-$\alpha$
processed helium, and other highly-processed material. 
This same process must also have involved the creation  of a cool 
supergiant which is currently contracting to become a white dwarf 
\citep{schoenberner86}. 

It was thought likely that EHe stars or RCB stars formed either
following a final helium shell flash (or late thermal pulse) in a
cooling white dwarf \citep{iben83,herwig00}, or following a merger
between a carbon-oxygen white dwarf and a helium white dwarf in a
close binary \citep{webbink84,saio02}. Consensus now strongly favours
the white dwarf merger origin, supported by evidence of evolution
timescales, pulsation masses, and surface abundances
\citep{saio02,pandey06,clayton07}. Efforts are now focused on 
securing the abundance measurements and on the question of whether
the merger is {\it cold} (no nucleosynthesis) or {\it hot}. For example,  
\citet{clayton07} argue that RCB overabundances of
oxygen in general and $^{18}{\rm O}$ in particular are produced
from nucleosynthesis during the merger, whilst \citet{pandey10} argue
semi-quantitatively that no additional nucleosynthesis is required to 
match the observed abundances of H, He, C, N, O and Ne in EHe stars. 

Evidently, we are far from an exhaustive understanding of (i) the evolution
and (ii) the subsequent merger of two white dwarfs in a close binary. In the
first instance, it is likely that the binary will have passed through
at least one prior phase of common-envelope evolution. In the second,
the merger of two white dwarfs involves the total destruction of the
less massive star and the assimilation of a subsequent hot disk into
the survivor. Both involve non-linear processes on dynamical
timescales. Consequently, the accuracy with which current surface
abundances may be used to infer past evolution may be legitimately
challenged.

The object of this paper is to clarify the surface abundances which
might arise under conservative assumptions for the merger of a helium
white dwarf (HeWD) with a carbon-oxygen white dwarf (COWD) and to
compare these with observed abundance anomalies. The question of
predicted birthrates and galactic distribution of double-white dwarf
mergers and their correlation with the observed distribution of EHe
and RCB stars will be addressed in a separate paper. Thus, the known situation
regarding surface abundances is reviewed in
\S~\ref{s:abunds}. Background assumptions and calculations relating to
the white dwarf merger model are discussed in \S~\ref{s:models},
particularly where these relate to the following.  The mixing model
used to infer merger surface abundances, together with the input from
detailed stellar evolution and nucleosynthesis calculations, is
described in \S~\ref{s:simple}. The inferred elemental abundances are
discussed in \S~\ref{s:elems}.  Finally, the sufficiency of the model
is discussed in terms of whether additional nucleosynthesis is
necessary. (\S~\ref{s:conc}).



\begin{table*}
  \vbox to220mm{\vfil Landscape table to go here:\\
  \caption{Observed abundances for EHe and RCB stars}
 \vfil}
 \label{t:obsabs}
\end{table*}

\begin{table}
  \caption{Notes to Table\,\ref{t:obsabs}}
 \label{t:obsrefs}
\begin{center}
\begin{tabular}{rl}
 A89 & \citet{anders89}   \\ 

 A00 & \citet{asplund00}  \\ 

 C05 & \citet{clayton05}  \\ 
 C06 & \citet{clayton06}  \\ 
 C07 & \citet{clayton07}  \\ 

 D98 & \citet{drilling98} \\


 J88 & \citet{jeffery88}  \\ 
 J92 & \citet{jeffery92}  \\ 
 J93a& \citet{jeffery93a} \\ 
 J93b& \citet{jeffery93b} \\ 
 J96 & \citet{jeffery96}  \\ 
 J97 & \citet{jeffery97}  \\ 
 J98 & \citet{jeffery98}  \\ 

 P01 & \citet{pandey01}   \\ 
 P06a& \citet{pandey06}   \\ 
 P06b& \citet{pandey06b}  \\ 
 P06c& \citet{pandey06c}  \\ 
 P08 & \citet{pandey08}   \\ 
 P10 & \citet{pandey10}   \\ 

\end{tabular}
\end{center}
\end{table}

\begin{figure*}
\begin{center}
\epsfig{file=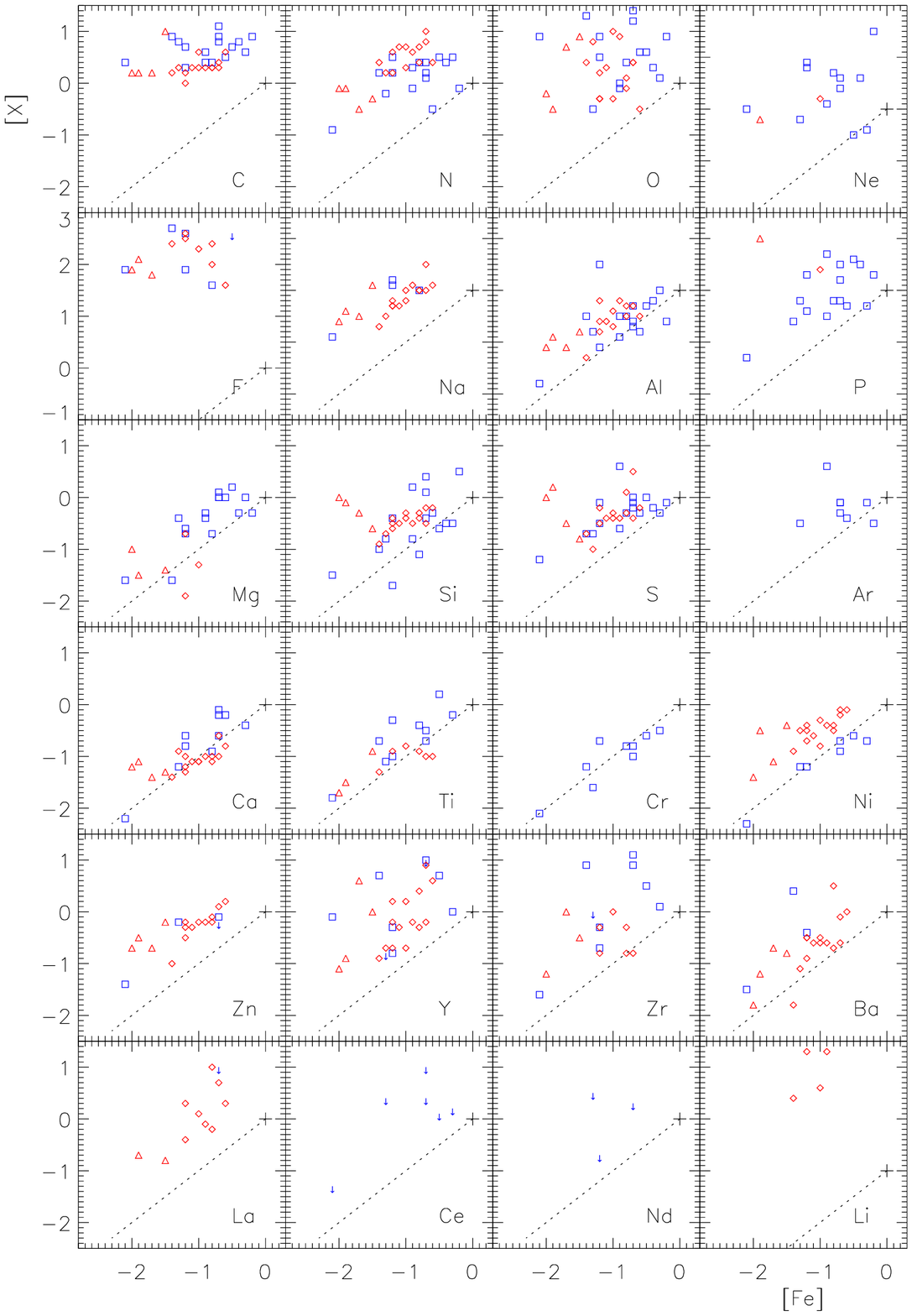,angle=0,width=140mm}
\caption[]{Observed surface abundances (log number relative to solar) 
  versus iron abundance (same units) for 
  extreme helium stars (blue squares), 
  majority RCB stars (red diamonds), 
  and minority RCB stars (red triangles). 
  Upper limits are shown as arrows.  
  The dotted line indicates a solar composition scaled to iron. 
  Note that the plots for Ne, F, and Li are offset vertically; a
  cross indicates $[Fe],[X]=0,0$ (solar abundance) in every case. 
}
\label{f:obsabs}
\end{center}
\end{figure*}

\section{The observed surface abundances of EHe and RCB stars}
\label{s:abunds}

Over the last two decades, much effort has been expended on the
accurate measurement of the surface abundances in EHe and RCB
stars. These have been derived primarily from the analysis of
high-resolution spectra using model atmospheres and theoretical line
profiles which assume local thermodynamic, hydrostatic and radiative
equilibrium (LTE).  Model atmospheres have been
computed using the codes MARCS \citep{asplund97a} for cool stars and
STERNE \citep{jeffery01b} for hot stars. RCB and very cool EHe
abundances are based on MARCS model atmospheres which 
use continuous opacities from the Opacity Project, and 
treat line opacities using opacity sampling.  Published
EHe abundances are based on STERNE models which use ATLAS6-type
continuous opacities and an opacity-distribution function computed
specifically for a hydrogen-poor carbon-rich mixture \citep{moller90}.

It is important to note that although the abundances derived are in
general consistent and not particularly sensitive to the model
atmosphere structure, there are likely to be exceptions for individual
elements. There are also improvements which could be made to
the model atmospheres, and would lead to greater overall confidence in the
derived abundances. For example, modern STERNE models include
Opacity-Project continuous opacities and opacity-sampling for lines
\citep{behara06}. These have not yet been used in any detailed fine
analysis of an EHe star, but do demonstrate sizable differences in
temperature structure and overall flux distribution from the earlier
models. For most RCB stars the major opacity source in the photosphere 
is carbon, but the most abundant species, helium, is not directly
observable. This has led to a ``carbon problem'' whereby it is
difficult to measure the carbon abundance unambiguously
\citep{asplund97a,pandey04b}. Since most EHe and RCB stars have low surface
gravities, departures from LTE may also be important, particularly for
the measurement of surface gravity and other quantities derived from
strong lines \citep{jeffery98,przybilla05,pandey10}.

For the present paper, published abundance data for EHe and RCB stars 
are collated and 
summarised in Tables~\ref{t:obsabs} and \ref{t:obsrefs}. 
Abundances are cited in logarithmic units\footnote{Conventionally,
  stellar abundances are given logarithmically by number, normalised such
  the logarithmic hydrogen abundance is equal to 12. This convention assumes that hydrogen 
  dominates the composition. In evolved mixtures, hydrogen may be
  vanishingly small, so the convention loses value. 
  The formalism given here preserves abundance values
  ($\epsilon_i$) of species unaffected, for example, by the 
  conversion of hydrogen to helium. } such that
\begin{equation}
\epsilon_i \equiv \log n_i + C, ~~~ \Sigma n_i = 1,
\end{equation}
where 
\begin{equation}
\log \Sigma n_i \mu_i + C = \log \Sigma n_{i \odot} \mu_i + C_{\odot}
\approx 12.15, \label{eq:cdef} \end{equation} 
and the normalisations are such that the logarithmic hydrogen abundance of the Sun
\begin{equation}
\epsilon_{\rm He \odot} \equiv 12.00. 
\end{equation}

Measurements have been rounded to 1 decimal place and given without
errors; the reader should refer to the papers cited for
more detail, but in general these are typically $\pm0.2-0.3$ dex.
For space
considerations, some elements measured in only a few stars 
have been omitted. Likewise, the HdC stars have been omitted, since 
iron abundances for these are not available. The EHe stars V652\,Her
\citep{jeffery99} and HD 144941 \citep{harrison97,jeffery97} have 
been excluded for a different reason. They are comparatively 
hydrogen-rich, nitrogen-rich and carbon-poor, suggesting a different
history. The measured iron abundance for DY\,Cen (5.0) is untypical of other 
elements \citep{jeffery93a}; unpublished data suggest a higher value. 
In this paper we assume an iron abundance for DY\,Cen 
scaled to that of aluminium, silicon and sulphur ($\epsilon_{\rm
  Fe}=7.3$).  

The emergence of overall patterns may be seen in Fig.~\ref{f:obsabs} where
each panel represents a different element, 
different symbols represent different groups of stars, 
$[X] \equiv \epsilon_i - \epsilon_{i\odot}$ represents the
(logarithmic) elemental abundance relative to the solar abundance, 
and [Fe] represents the iron abundance normalised in the same way. 
To understand this plot, consider that a star having the same composition as the Sun
would appear at the origin (0,0) in every panel. 
Stars with elemental abundances scaling exactly with the iron
abundance would lie on a straight line through the origin and having
gradient unity (as indicated by a broken line).  
Similar plots have been presented and discussed in detail by
\citet{asplund00} and \citet{pandey06}. In summary, their conclusions were as follows.

\subsection{Elements unaffected by evolution}

{\it Iron:} several elements appear to be representative of initial metallicity.
Fe may be adopted for spectroscopic convenience, and it is unlikely to
be affected by H and He burning and attendant nuclear
reactions. \citet{pandey06} find that Cr, Mn, and Ni vary in concert
with Fe, so that these may also be taken as proxies for the initial
metallicity.

\noindent {\it  $\alpha-$elements:} Mg, Si, S, and Ca and also Ti follow the
expected trend in which the abundance ratio $\alpha$/Fe varies with Fe
\citep{ryde04,goswami00} (with the possible exception of DY\,Cen: see above).

\noindent {\it  Aluminium:} abundances follow Fe with an
apparent offset of about 0.4 dex.
 
\noindent {\it  Argon:} in five out of seven EHes, Ar appears to have its
initial abundance.

\noindent {\it Nickel:} although Ni varies in concert with Fe in both EHes
and RCBs, there is a disconcerting offset of about 0.5 dex between the
two groups. At this juncture, one suspects a systematic error due to
the use of different ions or lines in the two groups of stars. The
difference serves as a reminder that caution must
be exercised with all abundance measurements discussed here.

\noindent {\it  Zinc:} like Ni and Al, Zn correlates well with Fe, with a
positive offset of $\approx0.8$ dex\footnote{At very low metallicity,
  Zn is thought to be enhanced by the s-process, but not at levels
  which concern us here.}. Again, one suspects a systematic error. 

\noindent {\it Minority RCBs:} \citet{Lambert94} identify a subset of four
RCBs which show lower Fe abundance and higher Si/Fe and S/Fe ratios
than the majority. These are indicated in Table~\ref{t:obsabs}.

\subsection{Elements affected by evolution}

{\it Hydrogen:} excluding DY\,Cen and V854\,Cen, the combined
sample of EHes and RCBs have H abundances $\log \epsilon_i$ in the
range 4 -– 8.

\noindent {\it Lithium:} a few RCBs are notably rich in lithium, which must
have been produced simultaneously with or subsequent to the process
which made these stars H-deficient \citep{asplund00}.

\noindent {\it Carbon:} excluding MV\,Sgr, the EHes show a mean carbon
abundance $\log \epsilon_i = 9.3$, and a range from 8.9 -- 9.7,
corresponding to a mean C/He ratio of 0.006 and a range from 0.003 to
0.010. The carbon abundance is more difficult to measure reliably in
RCBs; the mean indicated by Table~\ref{t:obsabs} is apparently lower
than in the EHes; this is probably a direct consequence of the 
carbon problem referred to above \citep{asplund97a}.

In RCB and HdC stars cool enough to show CO, the $^{12}{\rm
  C}/^{13}{\rm C}$ ratio is generally found to be greater than
100\footnote{the exception being V\,CrA with $^{12}{\rm C}/^{13}{\rm
    C}\approx 4$ \citep{rao08}} \citep{warner67,cottrell82}, confirming
a 3-$\alpha$ or helium-burning origin for the carbon excess.

\noindent {\it Nitrogen:} nitrogen is enriched in the great majority of EHes
and RCBs above that expected according to the Fe abundance
(Fig.~\ref{f:obsabs}). \citet{heber83} and subsequent authors point
out that the N abundances in general follow the trend expected by the
almost complete conversion of the initial C, N, and O to N via the
H-burning CN and ON cycles.  The exceptions are again DY Cen (very
N-rich for its Fe abundance), and LSS\,99 (very little N enrichment).

\noindent {\it Oxygen:} abundances relative to Fe range from underabundant
by more than 1 dex to overabundant by almost 2 dex. The stars fall
into two groups. Six EHes and a comparable number of RCBs with
[O/Fe]$\geq 1$ stand apart from the remainder which have an O abundance
closer to their initial value.  The O/N ratio for this remainder is
approximately constant at O/N $\approx$ 1 and independent of Fe.

Both groups present problems. There is no obvious means to produce
[O/Fe]$\geq 1$ (in most of these cases it is impossible to distinguish
$^{16}{\rm O}$ from $^{18}{\rm O}$). For the remainder, nearly all have an N
abundance indicating total conversion of initial C, N, and O to N via
the CNO cycles, so that an observed O abundance so close to the initial
abundance is unexpected. It could be partially accounted for by 
dredge-up of $^{16}{\rm O}$ from the CO-core during post-AGB
evolution, as has been suggested to explain high O abundances in
PG1159 stars \citep{werner06}. 

Another solution is suggested by the remarkable discovery that the
$^{18}{\rm O}/^{16}{\rm O}$ ratio in RCB stars (where observed) is
close to and sometimes greater than unity, a ratio many hundreds of
times higher than expected \citep{clayton05,clayton07}. In HdC
stars, the $^{18}{\rm O}/^{16}{\rm O}$ ratios are even higher
\citep{garcia09,garcia10}. However, there is as yet insufficient
evidence to indicate that the excess O is in the form of
$^{18}{\rm O}$ in {\it all} RCBs or in any EHes. 

Knowing that early-asymptotic giant branch stars possess a
high $^{18}{\rm O}/^{16}{\rm O}$ pocket in a thin layer of the
He-burning shell, \citet{warner67} speculated that 
under unusual circumstances a star might strip all of its envelope
material precisely down to this narrow layer, but 
\citet{clayton05} concludes this to be highly improbable. 

\noindent {\it Fluorine:} the discovery of very substantial quantities of
fluorine, first in several EHes and subsequently in most RCBs, was
also unexpected \citep{pandey06b,pandey08}. It appears to be uncorrelated
with Fe or O and is overabundant by 2 -- 4 dex. F is produced in
the He-intershell of an AGB star through a (complex) combination of $\alpha-$, 
n- and p-capture reactions \citep{lugaro04}, a conclusion 
confirmed by an observed correlation between C and F in the
atmospheres of AGB stars \citep{jorissen92,abia10}, and by
observations of post-AGB stars that show F at the level predicted to
be in the intershell \citep{werner09}. 

\noindent {\it Neon:} high overabundances derived from Ne{\sc i} lines for a
few intermediate temperature EHes were originally treated with scepticism --
non-LTE being a possible culprit. A similar overabundance measured from
Ne{\sc ii} lines in LS\,IV$+6^{\circ}2$ \citep{jeffery98} effectively
substantiated the Ne{\sc i} results in other stars. 
\citet{pandey10} have made a recent non-LTE analysis of
the neon abundances in three EHes, and confirmed a substantial 
overabundance approximately independent of the star's iron abundance. 
$^{22}$Ne is produced via two $\alpha$-captures on $^{14}$N, so should be
abundant in carbon-rich material derived from helium produced by the
CNO-process.

\noindent {\it Phosphorus:} an overabundance of P was first remarked in
BD$+10^{\circ}2179$ by \citet{hunger69}. This was discounted from
ultraviolet spectroscopy by \citet{heber83} and
\citet{pandey06}. Overabundances have been reported in several other
EHes by {\it inter alia}
\citet{kaufmann77,jeffery92,jeffery93a,jeffery93b}, where they are
systematically larger than in the sample studied by
\citet{pandey06}. Whether this represents a problem with $gf$-values
deserves further investigation. P can be produced through neutron
captures in an asymptotic giant branch star (of which more later). As
the observations stand, P overabundances may be a key diagnostic of
previous history.

\noindent {\it Heavy elements:} two EHes are severely enriched in Y and Zr:
V1920\,Cyg and LSE\,78 with overabundances of about a factor of 50
\citep{pandey04}. A third, PV\,Tel, is enriched by a factor of about
10. Five other stars for which measurements were possible
are considered to have their initial abundances of Y and Zr. Y and Zr
overabundances are attributed to contamination by s-process products.
The origin of these has not been identified.

Only upper limits are reported for rare earth elements La, Ce, and Nd,
all consistent with the observed abundances of Zr and Y.  The cool EHe
LS\,IV$-14^{\circ}109$ has a Ba abundance consistent with its initial
metallicity.

\subsection{Key Questions}

The surfaces of RCBs and EHes primarily exhibit 
CNO-processed helium. In addition, they show contamination by a
residue of hydrogen, by $3\alpha$-processed carbon, and by additional
$\alpha$-capture products. 

The primary challenge is to demonstrate a mechanism which will deliver
the observed mixture, or range of mixtures, at the stellar surface. 
This mechanism must also be able to explain large overabundances of 
Li, $^{18}$O, $^{19}$F, possibly P, and various s-process elements.


\section{Models for the evolution of CO+He WD mergers}
\label{s:models}

While evidence has accrued in favour of an origin involving the merger
of a carbon-oxygen white dwarf with a helium white dwarf, this has not
always been the favoured model. Valid questions include whether such
mergers can occur, with what frequency, and with what outcomes. 

\begin{table}
\caption{Galactic merger rates for double white dwarf binaries.}
\label{t:dd_merger}
\begin{tabular}{lccc}
\hline
Source & He+He & He+CO : CO+He & CO+CO \\
       & \multicolumn{3}{c}{${\rm yr^{-1}}$} \\
\hline
\citet{webbink84}$^1$
       & 2.9 & 1.9 & 1.2 \\
       & \multicolumn{3}{c}{$\times 10^{-11} {\rm pc^{-2} yr^{-1}}$} \\
\citet{iben96} & & 0.0023 &  \\
\citet{han98}$^2$ & 0.0112 &  0.0154 & 0.0044 \\
\citet{nelemans01} & & 0.0044 & \\
\citet{yu10} &  & 0.0027 &   \\
\hline
\end{tabular}
$^1$ It is not clear whether Webbink
  reports the double-degenerate birth rate or merger rate.\\
$^2$ Model Set 1: merger rate = birth rate assumes all DD's will merge
\end{table}

\subsection{The formation of CO+He binary white dwarfs} 
\label{s:dd_formation}

\citet{webbink84} first recognised that one consequence of close binary
evolution would be the formation and evolution of double white-dwarf binary
(DWD) systems that could ultimately merge.  

The evolution of a main-sequence star results in an
expansion that will bring it into contact with a sufficiently nearby
companion. Such contact may result in stable Roche lobe overflow, or
dynamical mass transfer, resulting in the formation and ejection 
of a common envelope. The outcomes depend on the binary mass ratio 
and on the structure of the larger star, and are diverse
\citep{iben85}. 
The remnants include double helium white dwarfs; carbon-oxygen plus
helium white dwarfs, and double carbon-oxygen white
dwarfs. \citet{webbink84} estimated birth rates for the formation of
each of these systems; similar estimates have formed one output of 
many subsequent binary-star population-synthesis studies of the
Galaxy.

Up to the mid 1980's, a criticism of theory was the absence of hard observational
evidence that short-period white-dwarf binaries actually do form, 
whether as a consequence of close-binary evolution or
otherwise. This problem was largely addressed by the 
discovery of significant numbers of such systems 
\citep{saffer88,marsh95a,marsh95b}. 
Further discoveries were made as a result of large-scale 
white dwarf surveys \citep{nap03,nelemans05,morales05}.

At present, there exists a qualitative agreement between observed 
DWD space densities and their predicted birth rates. 
The question, as it applies to DWD mergers, will be addressed in 
more detail elsewhere (Jeffery et al. in preparation).

\begin{figure}
\begin{center}
\epsfig{file=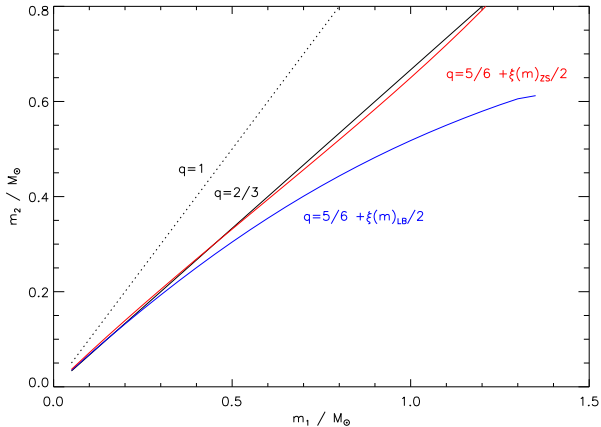,angle=0,width=85mm}
\caption[]{Stability limits for mass transfer in close double white
  dwarfs. Above the appropriate limit, mass transfer should be 
  dynamically unstable. The limits $\xi_{\rm LB}$ and $\xi_{\rm ZS}$
  refer to cold WD mass-radius relations due to \citet{lyndenbell01}
  and \citet{nelemans01} respectively. 
}
\label{f:qcrit}
\end{center}
\end{figure}

\subsection{Gravitational-wave radiation and dynamical mergers}
\label{s:gw_mergers}

There are two principles behind the idea that close-binary white
dwarfs will merge to form a single star. 

The first principle is that angular
momentum is removed from the binary by means of gravitational
radiation (GR) and that, within a Hubble timescale, the less massive
and consequently larger white dwarf will eventually fill its Roche
lobe and a phase of mass transfer will begin.
The timescale for orbital decay by GR is given by
\begin{equation} 
(\tau/{\rm y}) = 10^7 (P/{\rm h})^{8/3} \mu^{-1} (M/{\rm
M_{\odot}})^{-2/3}, 
\end{equation}
where $P$ is the orbital period, $M=m_1+m_2$ is the
total mass of the system and  $\mu$ is the reduced mass 
\citep{landau58,marsh95b}, indicating that DWD
systems with $P\simle$ a few hours will reach contact within a Hubble
time.

The second principle is that if the mass ratio
\begin{equation}
 q \equiv m_2/m_1 \geq q_{\rm crit} \equiv
 \frac{5}{6}+\frac{\xi(m_2)}{2}, 
\label{e:qcrit}
\end{equation}
the increase of radius due to the reduction of mass ($\xi(m) \equiv
{\rm d} \ln r / {\rm d} \ln m$) will exceed the increase in the Roche
radius caused by the transfer of angular momentum. Mass transfer
then becomes dynamically unstable and probably
causes the components to coalesce \citep{pringle75,tutukov79}. For this
paper we have adopted the mass-radius relation for a cold white dwarf
reported by \citet{lyndenbell01}, with $\beta=1.137$ and
$\mu_e=2.02$. Note that this relation, which is valid from
sub-planetary masses through to relativistic white dwarfs, gives 
$\xi(m)$ markedly different to that used by \citet{nelemans01}, which
lies very close to the classical non-relativistic 
$r \propto m^{-1/3}$ relation (Fig.~\ref{f:qcrit}).

For smaller $q$, mass transfer will be stable; but if the
mass-transfer rate exceeds the Eddington rate, the envelope of the
accretor will heat and expand leading to the possible formation of a
common envelope and which could also cause the stars to coalesce \citep{han99}. Only at 
the most extreme mass ratios will mass transfer be stable, possibly 
leading  to the formation of AM\,CVn systems (He+He WDBs)
\citep{nather81}. 
Significantly, {\it all} of the DWD systems for which mass ratios
could be measured \citep{maxted02} have $q>q_{\rm crit}$.

Using this information together with DWD birth rates, 
population synthesis studies indicate a DWD merger frequency for the Galaxy of
between $2.3 - 4.4 \, 10^{-3}$y$^{-1}$ \citep{iben90,nelemans01,yu10}. 
Other estimates are indicated in Table~\ref{t:dd_merger}, broken down
by binary type wherever possible. 

Of particular interest, of course, is the frequency of double
COWD mergers, since these are a possible  \citep{webbink84,iben84} but
arguable \citep{saio04,yoon05} source of Type
Ia supernovae (SN Ia). Other outcomes are more likely 
\citep{iben84,saio95,saio98}.  Double HeWD mergers may produce sdO 
\citep{webbink84} or sdB \citep{iben90} stars, which are ubiquitous in
old stellar populations \citep{brown01,busso05,lee05,rich05}.

\subsection{SPH simulations of the dynamical merger}
\label{s:sph_models}

Several simulations of the white dwarf merger process have been
attempted  using smoothed particle hydrodynamics
\citep{benz90,segretain97,guerrero04,yoon07,loren09}.
For CO+He WD mergers ({\it e.g.} 0.6+0.4 \Msolar), 
these demonstrate the total disruption of the low-mass WD
within roughly one orbital revolution ($\sim 90$s) and the conservation of
$\sim99$\% of its mass within a thick Keplerian disk. They also
demonstrate substantial prompt heating of the disrupted material, with
temperatures momentarily reaching several $10^9$\,K in the equatorial plane
\citep{guerrero04}.  
However, the temperatures are not extremely high,
the degeneracy of the disrupted material is rapidly lifted, and any
thermonuclear activity is ultimately quenched.

\subsection{Nucleosynthesis during a dynamical merger}
\label{s:hot_merger}

In previous discussion of the post-merger product, \citet{saio02} made
the simplifying assumption that {\it no} nucleosynthesis occurs during
the merger -- this is the {\it cold} merger approximation.

Where $^{12}{\rm C}$ and $^4{\rm He}$ mix at sufficiently high
temperatures, some thermonuclear activity will occur. Any $^{14}{\rm
  N}$ will also be briefly exposed to $\alpha$-burning. Using an
elegant one-zone model in which orbital energy is
converted to heat in a debris disk, \citet{clayton07} showed
that certain nuclear abundances could be demonstrably altered during the
merger. In particular, surplus $^{18}{\rm O}$ could be produced
through prompt nucleosynthesis of $^{14}{\rm N}$ and $^{4}{\rm
  He}$ from the debris of the helium white dwarf\footnote{The reactions involved are: $^{14}{\rm
    N}(\alpha,\gamma)^{18}{\rm F}(\beta^+)^{18}{\rm O}$}, 
without being subsequently destroyed by an additional $\alpha$-capture to form $^{22}{\rm Ne}$.

\citet{loren09} included a limited nuclear network in their dynamical
merger simulation, and reported nuclear yields for various DWD
progenitor combinations. Models which include some
nucleosynthesis during white dwarf destruction will be referred to 
as the {\it hot merger} approximation.

Since the physics of DWD mergers is of substantial wider interest for
the production of hot subdwarfs and SN Ia, the question of whether
mergers are {\it hot} or {\it cold} is particularly relevant. Rephrasing:
to what extent does nucleosynthesis occur as a direct consequence of
heat generated by orbital energy dissipated during the merger, and do
any nuclear products play a r\^ole in the subsequent evolution? In
particular, is a {\it hot merger} necessary to explain the high
$^{18}{\rm O}/^{16}{\rm O}$ ratio observed in RCB stars?

\subsection{Models of thermal and nuclear evolution after a merger}
\label{s:ev_models}

The evolution of a WD rapidly accreting helium  was
first considered long before the possibility of DWD mergers was widely recognised 
\citep{nomoto77}. This and subsequent calculations pursued the
evolution of the accretor through and beyond off-centre helium ignition
\citep{nomoto87,kawai87,kawai88,iben90,saio98,saio00,saio02}. 

Such models have been used to approximate evolution following a
dynamical merger by making some working assumptions.  These include the
less massive white dwarf being completely disrupted by the merger,
forming a Keplerian disk and subsequently being assimilated onto the
surface of the accretor. Assimilation has been assumed to be by
spherical accretion at half the Eddington rate\footnote{This rate was
  chosen to avoid a runaway explosion (for low accretion rates) and to
  satisfy energy conservation. Higher accretion rates would be
  possible if heat could be removed aspherically.} ($\approx 10^{-5}
{\rm M_{\odot} y^{-1}}$).  Accretion was switched off once a
pre-selected final mass was attained.

Conceptually, this approach is flawed. It implies that, following
shell-helium ignition and subsequent expansion, the reservoir of
material to be accreted ({\it i.e.}, the remnant of the disrupted
white dwarf) remains in a Keplerian disk deeply embedded within the
giant envelope. Although such a disk might survive, it runs counter to
the principle that viscous disks collapse on a much shorter timescale
\citep{lyndenbell74}. The disk is more likely to heat and expand to
form a high-entropy envelope in hydrostatic equilibrium
\citep{yoon07}.

In their simulations \citet{saio02} assumed all accreted material to
be deposited on the surface of the accretor, and to have a composition
defined by the mean composition of a helium white dwarf.  The chemical
structure of the accreting CO white dwarf was obtained by evolving a
star from the zero-age main sequence through to an appropriate point
on the white-dwarf cooling sequence. 

During post-merger evolution, there are two phases of convective
mixing. The first is bottom-up nuclear-driven convection immediately following
shell-helium ignition. In the \citet{saio00,saio02} models, this
occurs when the envelope mass is small, 
and does not produce very much mixing. Subsequent top-down
opacity-driven convection develops when the star becomes a giant, and
does not reach layers enriched by post-merger nucleosynthesis. The
models consequently show very little chemical enrichment at the 
surface from C, O or other nuclear products due to this
mixing. 

The relative absence of  $^{12}{\rm C}$ or $^{16}{\rm O}$ in the
\citet{saio02} COWD intershell contradicts the substantial 
enhancements predicted in the  AGB intershell ({\it e.g.}
\citet{herwig00}), possibly because chemical evolution through the
thermal-pulsing  AGB was not treated in sufficient detail.
Some additional merger sequences were therefore computed with 
enhanced  $\beta_{^{12}{\rm C}}=0.2$ and $\beta_{^{16}{\rm O}}=0.05$,
where $\beta_i$ represents the mass fraction of species $i$.

Consequently, \citet{saio02} argued that the dynamical 
merger would have to disrupt the outer layers of the COWD in order to
explain the observed EHe and RCB surface abundances of C. 
Scrutiny shows that the {\it simple recipe} adopted by
\citet{saio02}, and also by \citet{pandey01,pandey06,pandey10}, 
requires further refinement.

First, the COWD models would benefit from a more realistic abundance
distribution, particularly in the helium layer, which should have the
composition of the intershell region of the progenitor AGB star. 

Second, it was not appreciated that some  COWD models \citep[e.g.][]{saio02}
might contain a substantial pocket of  $^{18}{\rm O}$ at the interface
between the CO-core and the He intershell, as well as a reservoir 
of $^{22}{\rm Ne}$ in the CO-core. Thus, if the
outer layers of the CO-core are disrupted during the merger,
substantial  $^{18}{\rm O}$ and $^{22}{\rm Ne}$ will be dredged up in
addition to $^{12}{\rm C}$. 

This question is somewhat open. Not all post-AGB models show this $^{18}{\rm O}$
pocket. Depending on the intershell temperatures, $^{14}{\rm N}$ may
be completely destroyed by $\alpha$-capture to $^{22}{\rm Ne}$ {\it
  before} the helium-burning shell passes it into the CO core.

The object of the following sections is therefore to refine the
{\it simple recipe} used in previous discussions of EHe and RCB 
surface abundances, to incorporate a more realistic description of the  
chemistries involved, and hence to develop a more quantitative
framework in which to discuss white dwarf mergers as possible 
progenitors. 


\begin{figure*}
\begin{center}
\epsfig{file=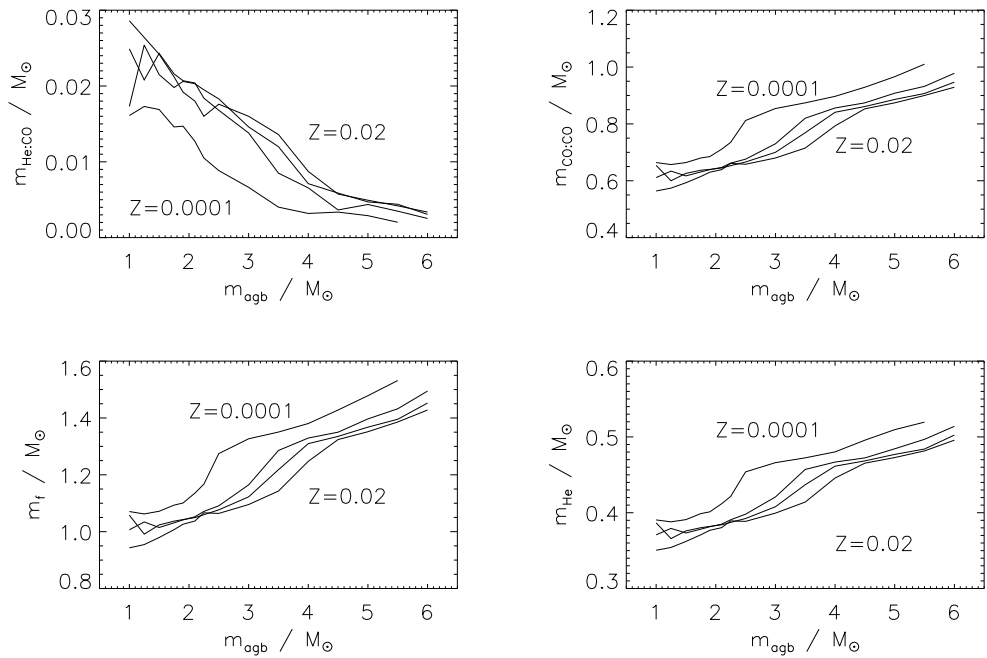,angle=0,width=140mm}
\caption[Merger component masses]{ Masses of the merger components as a function of the
    initial mass of the COWD progenitor ($m_{\rm agb}$) and of the 
    initial system metallicity ($Z = 0.0001, 0.004, 0.008$ and 0.02). 
    $m_{\rm He:CO}$ represents the final mass of the AGB intershell
    before the star contracts to become a COWD. 
    $m_{\rm CO:CO}$ represents the mass of the carbon-oxygen core at
    the same time. 
    $m_{\rm He}$ represents the mass of the HeWD secondary given by
    Eq.~\ref{eq:mhe}. 
    $m_{\rm f}$ represents the total mass of the product,
    assuming a conservative merger. 
}
\label{f:masses}
\end{center}
\end{figure*}

\

\section{The simple recipe}
\label{s:simple}

\subsection{Masses}

As is evident from the preceding summary, the processes by which two
white dwarfs may merge are far from straightforward. In addition, the
overall distribution of elements in both stars is a function of the
initial masses, metallicity and period in the original binary
system. To become a COWD, the ultimately more massive star must first
ascend the asymptotic giant branch, during which time the helium-rich
intershell will be processed by a series of thermal pulses, whilst the
hydrogen-rich envelope will be enriched by multiple convective
dredge-up episodes. To
create a close binary which will ultimately merge, the system must
pass through one or more common-envelope phases, producing an abrupt
change in the mass (at least) of one or both components. During the
merger, the less massive white dwarf will be completely disrupted to
form (a) a Keplerian disk and (b) a hot corona
\citep{yoon07,loren09}. During the disruption episode, temperatures
and densities may become high enough for nucleosynthesis. Shear mixing
between the disk, corona and the surface of the more massive white
dwarf seems inevitable, though how deep such mixing would be remains
to be explored. Subsequently, material from the disk and corona will
be accreted onto the surface of the more massive white dwarf; the
assumption is that it will be chemically homogenised. As this hot
material forces the star to expand and cool, surface convection zones
will develop from the surface.  Flash-driven convection will develop
following helium-shell ignition.  Most of these processes are poorly
understood. Where numerical models do exist, {\it e.g.} for the
dynamical phases \citep{yoon07,loren09}, or for the nuclear phases
\citep{saio02}, they are not yet joined up.

Until such time as they are, some simplifying assertions allow us to
make order-of-magnitude arguments. The principal of these is that all 
material from the helium white dwarf, and all hydrogen-rich
and helium-rich material from the carbon-oxygen white dwarf will be
fully mixed during the dynamical phase of the merger. From the point
of view of developing a recipe for calculating the chemical signature
of the merged product, this assertion gives us the first set of
parameters that will be required, namely the masses associated with
each layer of material to be mixed. We adopt the notation
 $m_{j:k}$ to represent the mass of layer $j$ of star $k$
\citep[see also][]{saio02}, thus:
\begin{description}
\item[$m_{\rm H:He}$:] the mass of the hydrogen-rich surface layer of the HeWD;
\item[$m_{\rm He:He}$:] the mass of the helium core of the HeWD;
\item[$m_{\rm H:CO}$:] the mass of the hydrogen-rich surface layer of
  the COWD; and
\item[$m_{\rm He:CO}$:] the mass of the helium shell of the COWD.
\end{description}
This principal assertion actually takes two forms. In the conservative case, all
material from both white dwarfs is included in the merged product. In
the non-conservative case, some mass may be lost under which
circumstance the above parameters represent the layer masses
which survive the merger. SPH calculations suggest that 
white-dwarf mergers are conservative \citep{guerrero04,loren09}. 

Without additional processing, none of these layers contains
sufficient carbon to account for the abundances
observed in EHe, RCB and HdC stars. As in previous applications of this
recipe, it is necessary to make a second assertion that
some material from the outer edge of the carbon-oxygen core of the
COWD has been somehow included into the mixture. For now, we define:
\begin{description}
\item[$m_{\rm CO:CO}$:] the mass of the carbon-oxygen core of the
  COWD; and
\end{description}
In addition, we have the total masses:
\begin{description}
\item[$m_{\rm He}$:] the total mass of the HeWD;
\item[$m_{\rm CO}$:] the total mass of the COWD;
\item[$m_{\rm f}$:] the final mass of the merged product.
\end{description}

In the conservative case, the number of variables is reduced since:
\begin{equation} m_{\rm f} = m_{\rm He} + m_{\rm CO}, \end{equation}
\begin{equation} m_{k} = \Sigma_j m_{j:k}; \quad j = {\rm H,He,CO}, k = {\rm He,CO}.
\end{equation}
Also, since helium (in general) dominates the final mixture and  
\begin{equation} m_{\rm He:He} >> m_{\rm He:CO} >> m_{\rm H:CO} \approx m_{\rm H:He},
\end{equation}
the observed hydrogen abundance naturally constrains the masses of the
hydrogen-rich layers:
\begin{equation} \beta_{\rm H}/\beta_{\rm He} \approx (m_{\rm H:CO} + m_{\rm H:He}) / m_{\rm He:He},
\end{equation}
where $\beta_i$ represents the abundance by mass fraction of species
$i$ ($\Sigma_i \beta_i \equiv 1$). 
Since in general EHes and RCBs show $\epsilon_{\rm H}/\epsilon_{\rm He}
\approx \beta_{\rm H}/(\beta_{\rm He}/4) < 10^{-4}$, we conclude
\begin{equation} m_{\rm H:CO} + m_{\rm H:He} < 10^{-4} m_{\rm He:He}. \end{equation}

We have attempted to use a similar argument to estimate the mass of
the  carbon-oxygen core to be included in the mixed layers 
($m_{\rm  mix:CO}$) by supposing that the observed C and O come
entirely from the COWD core; 
\begin{equation} (\beta_{\rm C}+\beta_{\rm O})/\beta_{\rm He} \approx m_{\rm mix:CO} / m_{\rm He:He}.
\end{equation}
It will be seen that this approach is too crude. 

With five masses to be adjusted in order to account for the abundances of 
hydrogen, helium, carbon, nitrogen and oxygen, the recipe appears 
under-constrained. Fortunately, stellar evolution 
theory provides additional information. 

For example, \citet{saio02} noted that in the conservative case for a 
0.6\Msolar\ carbon-oxygen white dwarf merging to form a 0.9\Msolar\ product,
the donor would likely have been 
predominantly helium with a mass $m_{\rm He:He}=0.3\Msolar$. 
On its surface would have been a hydrogen-rich envelope of mass $m_{\rm He:H}
\approx 10^{-3}-10^{-4}\Msolar$ \citep{driebe98}, 
reduced to this value by Roche lobe overflow during its first ascent of the
giant branch.

We now introduce the notion that a similar connection exists between
the carbon-oxygen white dwarf and the initial star in the binary system. 
Assuming that both stars evolve as single stars up to the point of
merger\footnote{The fully self-consistent approach would be to find the initial binary 
  $(m_1, m_2, P_{\rm orb})$ that will produce a close 
  white dwarf pair of appropriate dimensions and then to compute the 
  evolution of both components in detail, including their passage
  through any mass transfer or common-envelope phases. }, 
then models of stellar evolution through to the late
asymptotic giant branch \citep{karakas10} (Fig.~\ref{f:masses})
give values for $m_{\rm He:CO}$ and $m_{\rm CO:CO}$ in terms of 
\begin{description}
\item[$m_{\rm agb}$:] the initial mass of the star which becomes a COWD.
\end{description}
Such models also give a value for the mass of the AGB star hydrogen 
envelope,  but most of this will be substantially removed by stellar
winds and may form a planetary nebula before the star becomes a white dwarf.  

Establishing the mass of the COWD as a function of its progenitor mass,
the mass of the HeWD must lie below the minimum mass for
core helium ignition, approximately 0.48\Msolar,
and above the critical value for stable mass transfer (\S~\ref{s:gw_mergers}).
To restrict the number of free parameters in our model, we therefore set 
\begin{equation} m_{\rm He} = q_{\rm crit} m_{\rm CO}, 
\label{eq:mhe}
\end{equation}
the lower limit for dynamical mergers (Eq.~\ref{e:qcrit}). 
This automatically prescribes the ratio 
$m_{\rm He:He}:m_{\rm He:CO}$ for a given  $m_{\rm CO}$ and $Z$, 
and hence determines the dilution by the helium white dwarf
of elements produced in the AGB intershell. However, we note that more
massive HeWDs may exist, and that less massive HeWDs may merge 
as a result of a common-envelope phase. 

Thus, our {\it simple recipe} for predicting the surface abundances of
the product of a {\it cold, conservative} He+COWD merger 
now only requires $m_{\rm i:CO}$ and $Z$ as primary inputs, together
with a prescription for the composition of each component of the mixture.

\subsection{Composition}

We introduce the notation $\beta_{ijk}$ to refer to the mass fraction
of species $i$ in layer $j$ of component $k$. Generally $i \equiv z$, 
the atomic number. We currently consider the abundances of: 
$^1$H, $^4$He, 
$^{12}$C,  $^{13}$C,  $^{14}$N,  $^{16}$O,  $^{18}$O,  $^{19}$F,  $^{22}$Ne, 
$^{23}$Na, $^{24}$Mg, $^{27}$Al, $^{28}$Si, $^{31}$P,  $^{32}$S,  $^{40}$Ar,
$^{40}$Ca, $^{48}$Ti, $^{51}$V,  $^{52}$Cr, $^{55}$Mn, $^{56}$Fe, $^{59}$Co,
and $^{59}$Ni.  

Assuming the binary system was established with an initial metallicity
$Z$, the composition of the hydrogen-rich layers in both components
is then defined by a scaled solar composition ($\beta_{i {\odot}}$). 
We adopt:
\begin{align}
\beta_{i{\rm :H:He}} &=  \alpha_i(Z) \beta_{i {\odot}} . Z / Z_{\odot}, \qquad i > 2, \\
\beta_{\rm He:H:He} &= 0.28, \\
\beta_{\rm H:H:He} &= 1 - \Sigma_{i>1}\beta_{i{\rm :H:He}}, \\
\beta_{i{\rm:H:CO}} &=\beta_{i{\rm :H:He}}.
\end{align}

Generally, $\alpha_i = 1$, but in metal-poor environments ($Z<Z_{\odot}/10$) the abundances
of $^{16}$O,  $^{18}$O, Mg, Si, S, Ca, Ti and Mn are observed to exceed
the scaled solar value by as much as 0.5 dex
\citep{goswami00,ryde04}. $\alpha_i(Z)$ has been chosen accordingly.  
The last relation implies that dredge-up into the H-envelope 
has been ignored, especially dredge-up on the AGB, but also 
dredge-up prior to the AGB.  
This is justified because $m_{\rm H:CO}$ is so small
that, apart from hydrogen, this layer makes a negligible contribution
to the merger composition.

\subsubsection{HeWD core} 

The composition of the HeWD core is assumed to have been
produced by CNO-cycle hydrogen burning. This converts practically 
all of the carbon and oxygen (depending on temperature) to $^{14}N$. 
To allow for incomplete CNO cycling in low-mass stars, 
we introduce the branching ratio $f_{\rm CNO}$ between 
the full CNO cycle and the CN cycle. The composition is then given by:
\begin{align}
\beta_{i{\rm :He:He}} &=  \beta_{i{\rm :H:He}}, {\rm except \ldots}\\
\beta_{\rm ^{12}C:He:He} &= 0, \\
\beta_{\rm ^{13}C:He:He} &= 0, \\
\beta_{\rm ^{14}N:He:He} &= \Sigma_{\rm C,N} \beta_{i{\rm :H:He}} + f_{\rm CNO} \Sigma_{\rm O} \beta_{i{\rm :H:He}}, \\
\beta_{\rm ^{16}O:He:He} &= (1 - f_{\rm CNO}) \beta_{\rm ^{16}O:H:He}, \\
\beta_{\rm ^{18}O:He:He} &= (1 - f_{\rm CNO}) \beta_{\rm ^{18}O:H:He}, \\
\beta_{\rm H:He:He} &= 0, \\
\beta_{\rm He:He:He} &= 1 - \Sigma_{i\neq2}\beta_{i{\rm :He:He}}. 
\end{align}
$\Sigma_{i}$ represents a sum over all isotopes of species $i$. 
So far, we have always used $f_{\rm CNO}=1$, implying the full CNO cycle.  

\subsubsection{COWD shell}

The helium-rich layer of the COWD corresponds to the intershell of the
progenitor AGB star. The composition of this layer is the most
interesting of all since it contains a combination of CNO-cycled
helium, various $\alpha$-capture products, and the products of a
nuclear network which includes s-process neutron-capture 
products. The yield of each isotope from these processes is a
sensitive function of the initial abundances and of the temperatures 
and densities throughout successive thermal-pulse cycles of the AGB
star. Given the importance of this layer to the final composition of
the merged product, we have adopted intershell compositions from a grid
of full AGB-star evolution calculations \citep{karakas10}. These 
provide fractional abundances for  
$^4$He, 
$^{12}$C,  $^{16}$O,  $^{17}$O,  $^{18}$O,  $^{19}$F,  
$^{20}$Ne, $^{21}$Ne, $^{22}$Ne,
$^{23}$Na, $^{24}$Mg, $^{25}$Mg, $^{26}$Mg,
$^{27}$Al, $^{28}$Si, $^{29}$Si, $^{30}$Si,
$^{31}$P,  $^{32}$S,  $^{33}$S, and $^{34}$S and 
define $\beta_{\rm i:He:CO}$ for these species. In our recipe, 
the isotopes of Ne, Mg, Si and S, are combined, since these 
are not resolved observationally.
$^{13}$C and $^{14}$N are destroyed in the intershell, with  $^{14}$N
being converted to $^{22}$Ne. 
Thus:
\begin{align}
\beta_{\rm H:He:CO} &=  0, \\
\beta_{\rm ^{13}C:He:CO} &= 0, \\
\beta_{\rm ^{14}N:He:CO} &= 0.
\end{align}
The AGB model grid of \citet{karakas10} provides 
information about light-element abundances, core and shell masses in
post-AGB stars over a large range of $m_{\rm agb}$ and $Z$, but lacks
detailed information for s-process elements. 
Consistent calculations for Zn, Y, Zr, Ba, La, Ce and Nd are available 
for one model with $m_{\rm agb}=2 \Msolar, Z=0.0001$ (Lugaro et
al. 2011, in preparation) and one with 
$m_{\rm agb}=3 \Msolar, Z=0.02$. 
We have used these as indicative of the range of s-process
yields in AGB stars of intermediate initial mass. 

\subsubsection{COWD core boundary}

\label{s:model:mix}

The outer layers of the carbon-oxygen core of the COWD will obviously
lack hydrogen and helium and be dominated by $^{12}$C and $^{16}$O
from the 3-$\alpha$ and $^{12}{\rm C}(\alpha,\gamma)^{16}{\rm O}$ reactions.  

An initial approach was to adopt a factor $f_{\rm CO}\equiv \beta_{\rm
  ^{12}C} / (\beta_{\rm ^{12}C}+\beta_{\rm ^{16}O}) = 0.8$ based on
typical abundances for COWD cores from contemporary
computations. Since the outer layers of the core will be those which
most recently exited the base of the intershell, the abundances of
most other elements can be set equal to those of the intershell:

\begin{align}
\beta_{\rm H:CO:CO} &=  0, \\
\beta_{\rm He:CO:CO} &=  0, \\
\beta_{\rm ^{16}O:CO:CO} &= (1 - f_{\rm CO}) (\beta_{\rm He:He:CO} +
     \beta_{\rm ^{12}C:He:CO} + \beta_{\rm ^{16}O:He:CO})  \\
\beta_{\rm ^{12}C:CO:CO} &= f_{\rm CO} (\beta_{\rm He:He:CO} +
     \beta_{\rm ^{12}C:He:CO} + \beta_{\rm ^{16}O:He:CO})   \\
\beta_{i:{\rm CO:CO}} &=  \beta_{i:{\rm He:CO}}, \quad i\neq1,2,6,8
\end{align}

Apart from the introduction of realistic intershell abundances, 
this approach follows that discussed by \citet{saio02}, which 
failed to explain the oxygen abundances without recourse to 
an {\it ad hoc} argument. Following discoveries of a large 
$^{18}$O excess in several RCB and HdC stars
\citep{clayton07,garcia09}, the models for COWDs used 
by \citet{saio02} were re-examined and found to contain a 
substantial pocket of  $^{18}{\rm O}$   
at the interface between the CO-core and the He intershell, 
as well as a reservoir of $^{22}{\rm Ne}$ in the CO-core.

\begin{figure}
\begin{center}
\epsfig{file=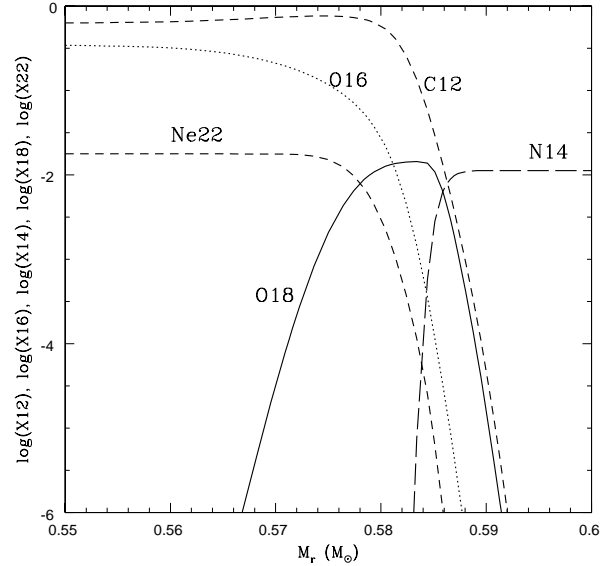,angle=0,width=85mm}
\caption[]{The distribution of various elements by mass fraction in
  the outer layers of a 0.6 M$_{\odot}$ post-AGB star evolving
  towards the ``knee'' of the white dwarf sequence. }
\label{f:cd}
\end{center}
\end{figure}

The effective mass and composition of this pocket, as it would
contribute to the merged white dwarf is not well constrained by the
models we have available. For example, as the star leaves
the AGB, the mass of this $^{18}{\rm O}$ pocket may be some
$3.10^{-4}$\Msolar. Subsequent steady He-burning in the post-AGB phase
(Fig.~\ref{f:cd}) 
produces an $^{18}{\rm O}$ pocket in the pre-merger white dwarf of 
 $\approx 0.008$ \Msolar (FWHM) having a mean $^{18}$O abundance 
$\beta_{^{18}O:CO} \rangle \approx 0.01$.

In the models of \citet{saio02}, this pocket is destroyed 
when the pocket is reheated by the post-merger accretion-driven helium 
flash; a new  but smaller pocket is established at the outer edge of the newly
established He-burning shell. However, if the  $^{18}$O pocket or material from
deeper in the CO core is mixed during the merger, it may survive. 
Recalling that there is insufficient carbon in the post-AGB 
intershell to account for all the carbon observed in EHe and RCB
stars, it is highly plausible that any carbon mixed into 
the merger-product envelope from the CO core  will be accompanied by
$^{18}$O. This will be diluted by additional $^{12}$C and $^{16}$O 
if the mixing penetrates significantly beyond the $^{18}$O pocket. 

We have therefore incorporated the outer layers of the CO core more 
realistically. The distribution with mass of \iso{4}{He}, \iso{12}{C}, 
\iso{14}{N}, \iso{16}{O}, \iso{18}{O} and \iso{22}{Ne} is assumed to resemble that 
of the $0.60 \Msolar$ pre white dwarf shown in Fig.~\ref{f:cd}, 
where a core mass of $0.58 \Msolar$ is defined 
by the point where the carbon and helium abundances are approximately equal.
This composition distribution can be applied to COWD cores of
different mass ($m_{\rm CO:CO}$) by scaling its thickness inversely 
as $(0.58/m_{\rm CO:CO})^4$ (this scaling also approximately reproduces the $m_{\rm He:CO}-m_{\rm CO:CO}$ relation
implied by Fig.~\ref{f:masses}). 

The abundances of \iso{14}{N}, \iso{18}{O} and \iso{22}{Ne} may be 
scaled with metallicity $Z$, since  \iso{14}{N} derives mainly from 
the initial metallicity, and  \iso{18}{O} and \iso{22}{Ne} are mainly 
formed from $\alpha$  captures on \iso{14}{N}. 

To explore the consequences of varying the \iso{12}{C}:\iso{16}{O}
ratio in the outer core, the factor $f_{\rm CO}$ can be used to force
a rescaling of these two species. The model shown in Fig.~\ref{f:cd}
has $f_{\rm CO}=0.8$, but our AGB models suggest $f_{\rm CO}=0.5$, 
probably reflecting differences in the adopted $\iso{12}{C}(\alpha,\gamma)\iso{16}{O}$ rate. 

A parameter $\alpha_{\rm mix}$ represents the mass within the boundary layer which is mixed 
into the merged envelope. Formally, if $\beta(m_r)$ represents the distribution of mass fraction
with mass inside the star ($m_r$), we compute
\begin{equation}
\beta_{i\rm :CO:CO} = \int_{m_i}^{m_o} \beta_{i\rm :CO:CO}(m_r) d m_r
/ \int_{m_i}^{m_o} d m_r
\end{equation}
for $i$ corresponding to \iso{12}{C}, \iso{14}{N}, \iso{16}{O},
\iso{18}{O} and \iso{22}{Ne}. 
The mass limits are given by 
\begin{align}
m_{\rm o} &= m_{\rm co:co} + m_{\rm sh}, \\ 
m_{\rm i} &= m_{\rm o} - \alpha_{\rm mix} m_{\rm sh}, \\
m_{\rm sh}&= (0.12/m_{\rm co:co})^4.
\end{align}
The shell mass $m_{\rm sh}$ characterises the scaled shell thickness 
corresponding to 0.12\,\Msolar\ in the 0.60\,\Msolar\ pre-WD model. 

Thus $\alpha_{\rm mix}=0$ means that no
carbon-enriched material from the boundary layer is mixed. 
$\alpha_{\rm mix}=1$ implies that all material down to the point where 
the carbon and helium abundances are equal is mixed. 
 $\alpha_{\rm mix}=2$ means that the mixed layer reaches the
region where \iso{22}{Ne} and \iso{16}{O} are abundant. 

Other $\beta_{i\rm :CO:CO}$ are as given previously, except
\begin{equation}
\beta_{\rm He:CO:CO} = 1 - \Sigma_i \beta_{i\rm :CO:CO},
\end{equation}
since this layer includes some helium from the base of the helium
layer. The layer masses $m_{jk}$ are adjusted to take the blurring of the
carbon-helium layer boundary into account. 

\subsubsection{Final abundances}

The ingredients of our model thus comprise five layers of
material with masses $m_{jk}$, each having a representative
composition $\beta_{ijk}$ defined by current stellar evolution theory. 
Assuming our primary assertion that all of these layers are fully
mixed during the merger, and that no further nucleosynthesis affects
the apparent surface composition of the merged product, then the
latter is simply represented by 
\begin{equation}
\beta_{i} = \Sigma_{jk} m_{jk} \beta_{ijk} / \Sigma_{jk} m_{jk} 
\end{equation}

For comparison with observation, these abundances can be transformed
to units more familiar in observational studies. Recall that mass
fraction $\beta$ is defined in terms of number fraction $n$:
\begin{equation}
\beta_{i} = n_i \mu_i / \Sigma_i n_i \mu_i
\end{equation}
whence 
\begin{align}
\epsilon_i \equiv \log n_i & = \log \frac{\beta_i}{\mu_i} + \log
\Sigma_i \mu_i n_i , \\
& = \log \frac{\beta_i}{\mu_i} + C' \\
C' & = \log \Sigma_i n_{i \odot} \mu_i - C \\
[X] & = \epsilon_{i} - \epsilon_{i \odot}, 
\end{align}
where $n_i$ are the relative abundances of species $i$ by number
($\Sigma n_i = 1$), and $C$ is defined by Eq.~\ref{eq:cdef}.


\begin{figure*}
\begin{center}
\epsfig{file=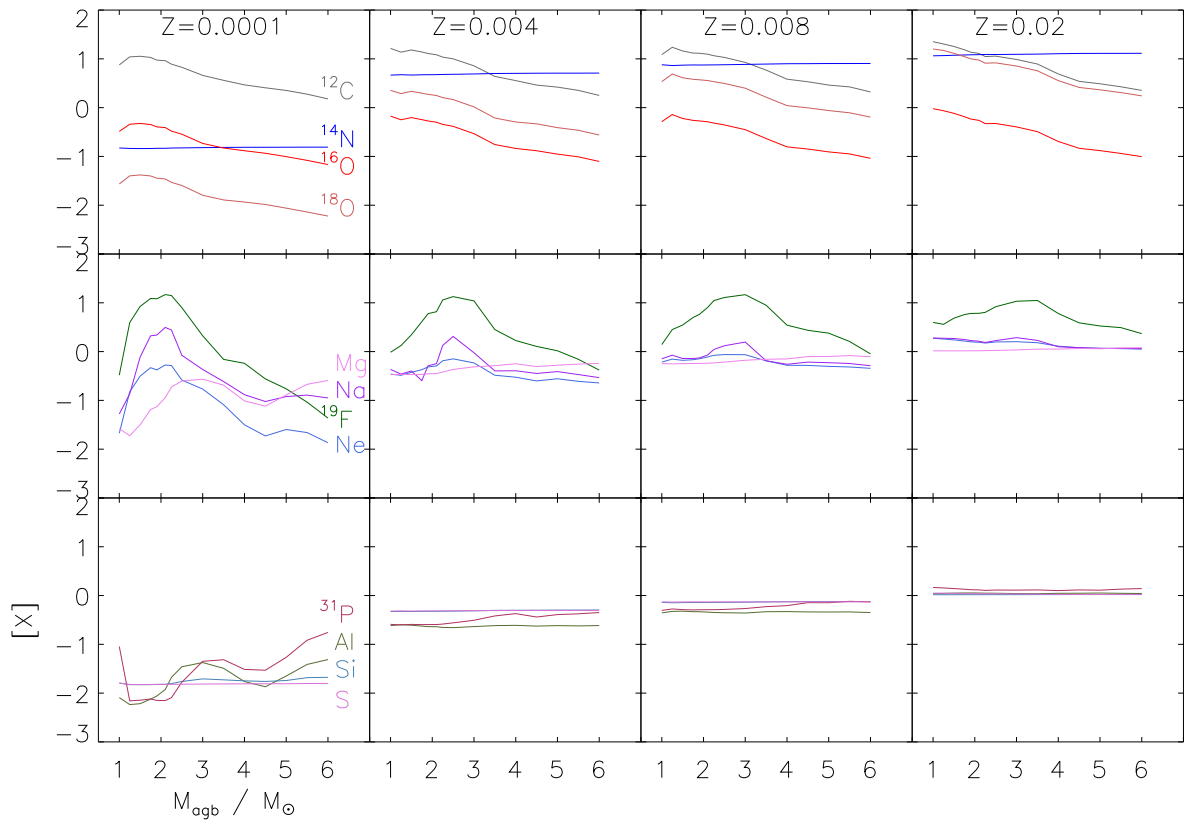,angle=0,width=180mm}
\caption[]{Surface abundances (log number relative to solar) predicted
  from our simple recipe for a merged white dwarf as a function of
  initial mass for the COWD ($m_{\rm agb}$) for four values of initial
  metallicity ($Z$). Elements and isotopes are coded by colour (shade of
  grey) and  labelled in the left column.}
\label{f:mbwd_yieldsbym}
\end{center}
\end{figure*}

\begin{figure*}
\begin{center}
\epsfig{file=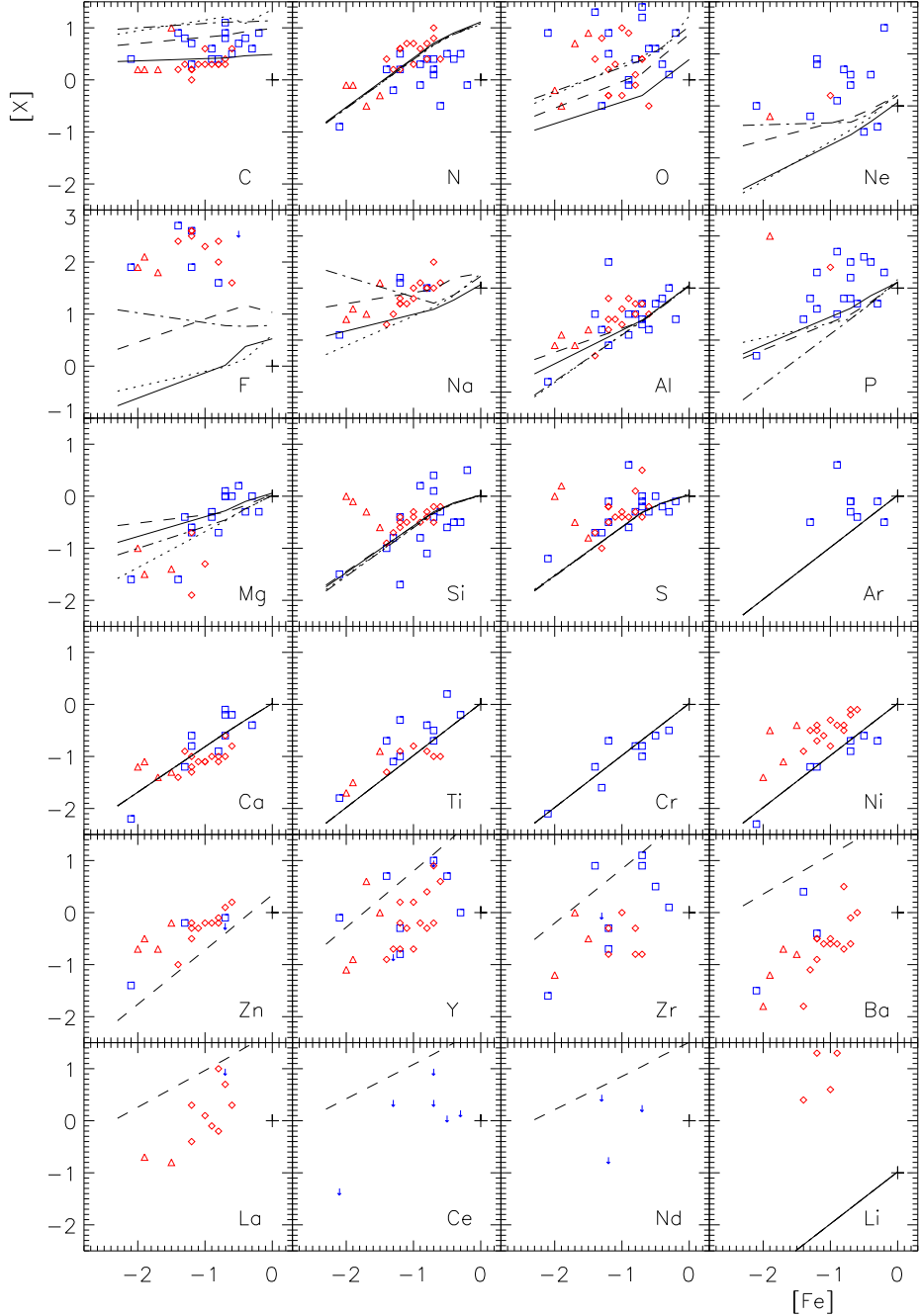,angle=0,width=140mm}
\caption[]{As Fig.~\ref{f:obsabs}. The over-plotted lines 
  represent the surface abundances predicted from a cold CO+He white dwarf
  merger. Predictions for four initial masses are shown:
  $m_{\rm agb}/{\rm M_{\odot}}=1$ (dotted), 1.9 (dash-dot), 3 (dashed) 
  and 5 (solid). Two or more lines are coincident in several panels,
  especially where only a single solid line appears. The very provisional
  result for s-process elements (see \S 4.2.2) is represented
  by a single dashed line. In this simulation, most 
  of the the carbon and oxygen are 
  dredged from the carbon/helium boundary layer at the top of CO core
  ($\alpha_{\rm mix}=3, f_{\rm CO}=0.8$). 
}
\label{f:modvobs}
\end{center}
\end{figure*}

\begin{figure*}
\begin{center}
\epsfig{file=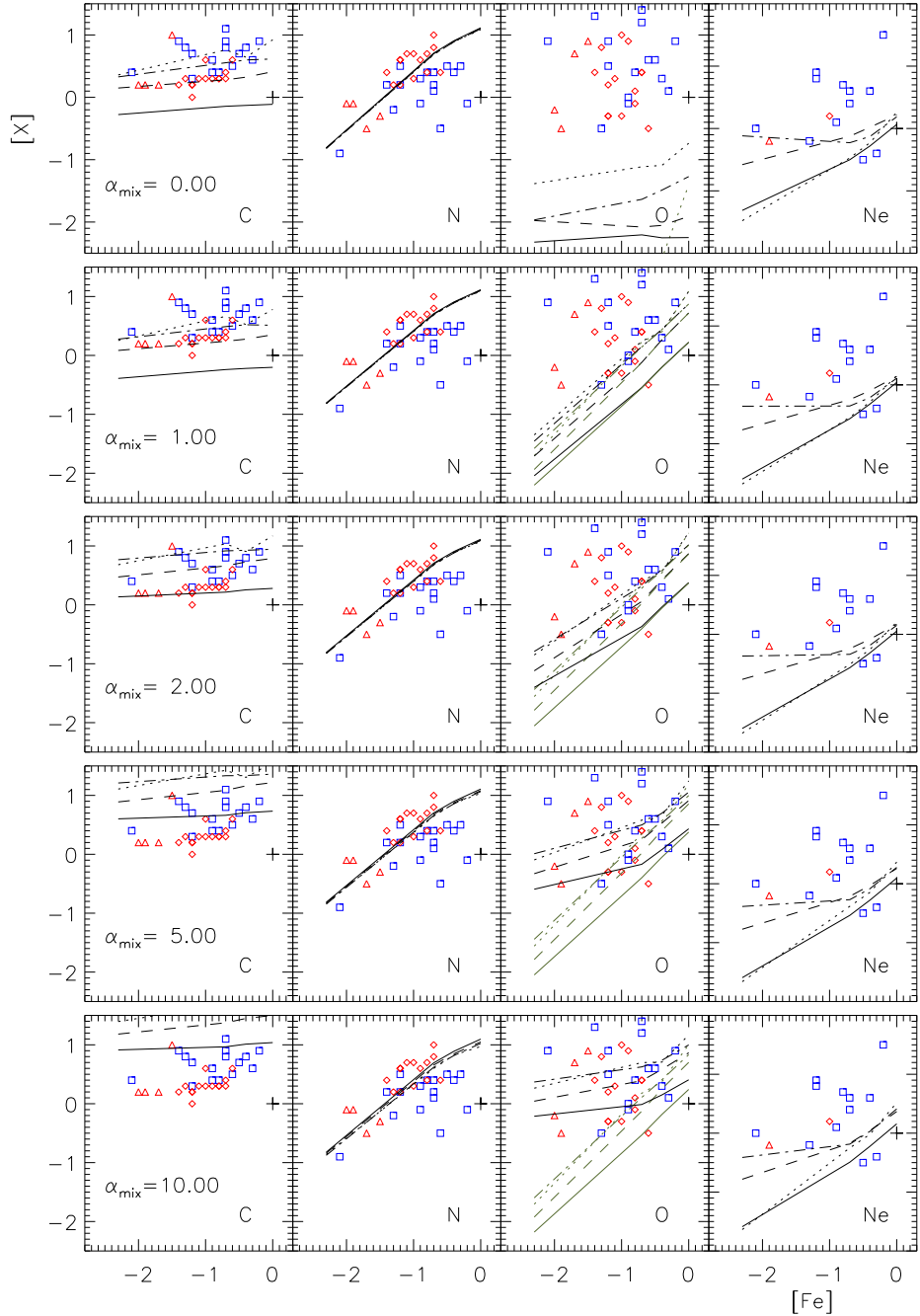,angle=0,width=140mm}
\caption[]{As the top 4 panels of Fig.~\ref{f:obsabs} repeated with 
  different values for the core-shell boundary mixing parameter
  $\alpha_{\rm mix}$. The top row represents the inclusion of {\it no}
  helium-depleted material from the boundary layer. The second row
  represents mixing down to a point where carbon and helium
  abundances are equal. The bottom three rows represent 
  increasingly deep mixing into material stratified as in
  Fig.~\ref{f:cd}; $\alpha_{\rm mix}=1$ includes 50\% of the
  \iso{18}{O} pocket. The contribution of \iso{18}{O} to the total
  oxygen abundance is shown in olive (grey). 
}
\label{f:alphamix}
\end{center}
\end{figure*}

\begin{figure*}
\begin{center}
\epsfig{file=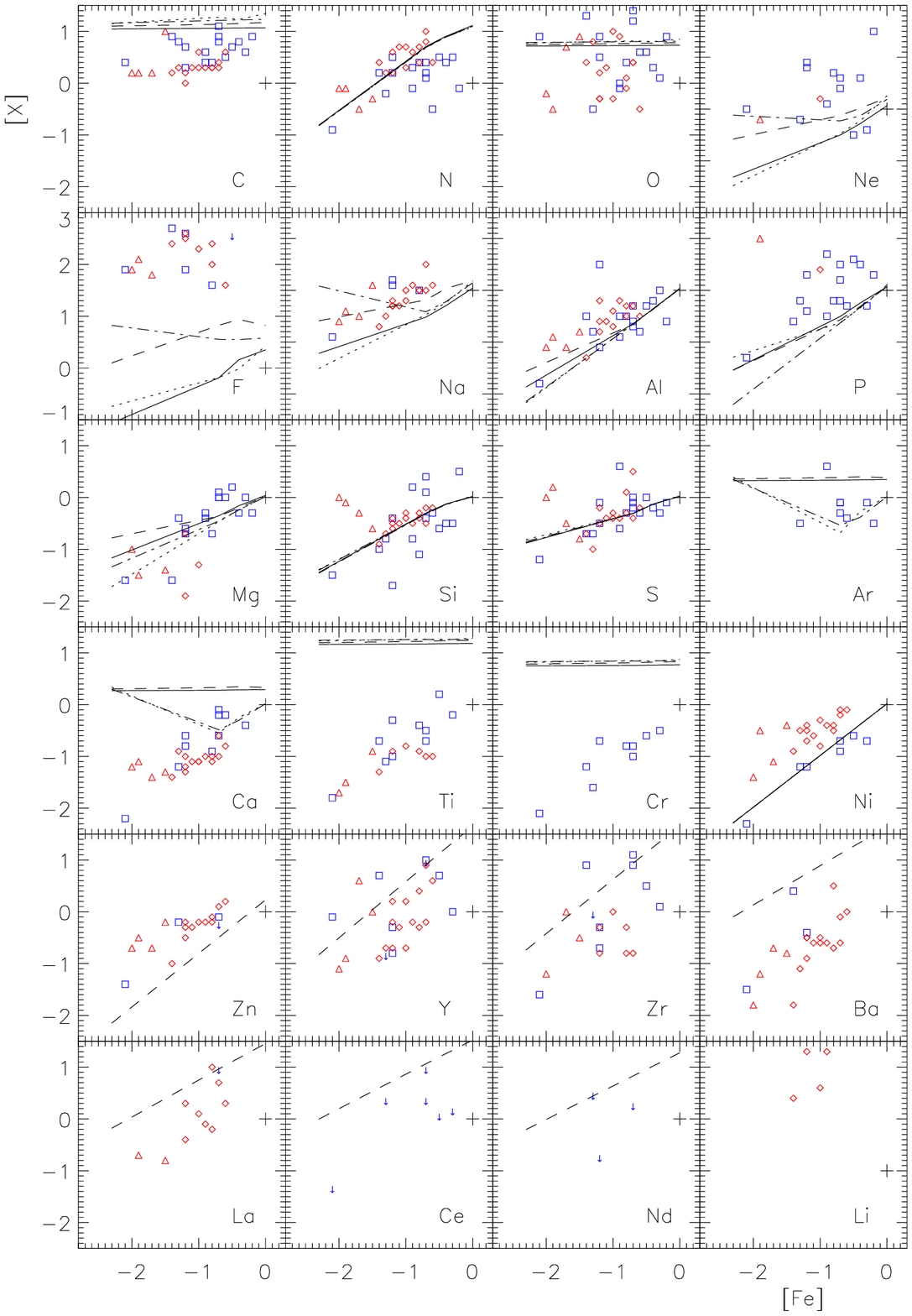,angle=0,width=140mm}
\caption[]{As Fig.~\ref{f:modvobs} except that the over-plotted lines 
  represent the surface abundances predicted from a hot CO+He white dwarf
  merger and include no mixing with the CO
  core ($\alpha_{\rm mix}=0$). Nucleosynthesis in the disrupted He-WD during the merger 
  is interpolated between results for a $0.3+0.5 {\rm M_{\odot}}$ 
  and a $0.4+0.8 {\rm M_{\odot}}$ merger \citep{loren09}. The
  predictions for Ti and Cr are based on the model for a $0.3+0.5 {\rm
  M_{\odot}} $ merger only.  
}
\label{f:hotvobs}
\end{center}
\end{figure*}

\subsection{Nucleosynthesis during a hot merger}

In the {\it cold} merger model, the chemical composition of the HeWD 
is assumed to be unchanged during a merger with a COWD. However, SPH
calculations indicate that some of this material may briefly reach
temperatures of $6\times10^8$K or more \citep{guerrero04,yoon07,loren09}, and
that some nucleosynthesis of $\alpha$-rich material will occur.  
\citet{yoon07} and \citet{loren09} demonstrate how the disrupted
material forms a relatively unprocessed disk containing slightly more than
half of the HeWD, and a heavily processed corona containing the
remainder. How the disk and corona are subsequently assimilated into
the merged star, and what mixing processes occur, is not yet
obvious. The simplest assumption is that turbulent mixing during the
merger, and nuclear- and surface-driven convection following stable
helium-shell ignition will completely mix both disk and corona with 
material from the intershell of the COWD. 

This process can be incorporated into our model. \citet{loren09}
(Tables 1 and 2) give sample masses and chemistries for the disk and 
corona in the cases of a $0.3+0.5 {\rm M_{\odot}}$ and a 
$0.4+0.8 {\rm M_{\odot}}$ He+CO WD merger. 
Using $m_{\rm He:He}$ from Eq.~\ref{eq:mhe} and interpolating
we obtain $m_{\rm disk:He}$ and  $m_{\rm corona:He}$ for $0.3<m_{\rm
  He:He}<0.4$ (we do not extrapolate), and also chemistries for the
same material.  \citet{loren09} compute models starting with a pure
helium mixture for the HeWD and a carbon-oxygen mixture for the 
CO white dwarf, so we arbitrarily impose a Z-dependent 
lower limit on individual abundances in the disk and corona 
as given by the prescription for the {\it cold} merger. We note 
that  contamination by a small amount of hydrogen and other metals will
profoundly affect the predicted {\it hot merger} nucleosynthesis, but
we have no data on how this might affect current results.

\section{Elemental yields}
\label{s:elems}

\subsection{Cold merger}
\label{s:elems:cold} 

The predicted surface composition following the {\it cold} merger 
of a He+CO WD with some core mixing is illustrated in 
Figs.~\ref{f:mbwd_yieldsbym}, \ref{f:modvobs}, and \ref{f:alphamix}.

Fig.~\ref{f:mbwd_yieldsbym} shows the surface
abundance of each species relative to the solar surface abundance as a
function of $m_{\rm agb}$ (the progenitor mass of the AGB star) for
each of four initial metallicities $Z$. This figure demonstrates that 
several species show high yields roughly independent of $m_{\rm
  agb}$. \iso{12}{C} and \iso{18}{O} show uniformly high yields as defined by the
core mixing parameters. The \iso{16}{O} yield is proportional to the
\iso{12}{C} yield; effectively determined by the
\iso{16}{O}/\iso{12}{C} ratio in the outer edge of the core. 
The \iso{14}{N} yield is proportional to $Z$ as expected; it is
determined by the initial CNO abundance. 
Note how the yields of $^{19}$F, Na and Ne are strongly peaked in the 
interval $1.5 < m_{\rm agb}/\Msolar < 3$, whilst $^{31}$P and Al are
only significantly enhanced ($\geq 0.5$ dex) toward higher masses 
($m_{\rm agb}>2\,\Msolar$). These yields reflect
the AGB intershell nucleosynthesis. Meanwhile $^{12}$C, $^{14}$N and
$^{18}$O are enhanced almost uniformly with $m_{\rm agb}$ and $Z$,
effectively as the model was designed to deliver.

Fig.~\ref{f:modvobs}  shows these same data rearranged as a 
function of initial metallicity ${\rm [Fe]}\equiv \log Z /
Z_{\odot}$ for four representative masses $m_{\rm agb}=1, 1.9, 3$ and
$5\Msolar$. They are plotted together with the observed abundance data
in exactly the same way as in Fig.~\ref{f:obsabs}. 
Since we have
only used two AGB models for the s-process elements, a single broken
line represents the provisional prediction for $m_{\rm
  agb}\approx3$\,\Msolar. 
Several correlations are noteworthy.

\subsection{Cold merger: individual elements}
\label{s:cold:els}

\noindent {\it Carbon:} the model carbon enhancement was chosen to
match the abundances measured in EHe stars by adjusting the degree of
core mixing.  The EHe carbon measurements are probably more
representative than the RCB measurements because of the ``carbon
problem'' referred to earlier. The EHe star data show a significant
scatter. Allowing $\alpha_{\rm mix}$ to vary by $\pm1$ 
gives results still broadly consistent with the observed carbon 
(and oxygen) abundances. 

\noindent {\it Nitrogen:} a generally excellent correlation with the
observations, supporting the basic assumption that the surfaces of EHe
and RCB stars are primarily CNO-processed helium.

\noindent {\it Oxygen:} a modest \iso{18}{O} pocket at the
helium-carbon boundary in the CO white dwarf progenitor may explain
the most extreme oxygen abundances seen in metal-rich EHe and RCB
stars, 
providing this layer is mixed during or immediately after the merger.  Note,
however, that oxygen isotope ratios are known for only a few RCB stars
and for no EHe stars, that \iso{18}{O} may be destroyed by high
temperatures during the merger. An alternative is that the 
\iso{16}{O}:\iso{12}{C} ratio in the outer core is approximately unity
or more.  

\noindent {\it Neon:} intermediate-mass models ($1.9, 3 \Msolar$)
show significant enhancements of neon, formed primarily in the
intershell of the AGB precursor. These match a few of the 
observed neon abundances, but the very high abundances 
measured in at least seven EHes are not yet explained by this model. 

\noindent {\it Fluorine:} as in the case of neon, a significant
\iso{19}{F} excess is generated in the intershell of the 
intermediate-mass models ($1.9, 3 \Msolar$), suggesting a likely
source for the observed excess. However, the predictions remain
$\approx 1$ dex smaller than the measurements. However, \iso{18}{O}
from the C/He boundary in the COWD provides an ample reservoir for the 
prompt production of \iso{19}{F} if it is sufficiently heated,
together with protons, during the merger; thus the model prediction
represents a strict lower limit.  

\noindent {\it Sodium, Aluminium, and Magnesium:} 
the recipe predicts some enhancement of these light elements over 
their initial values, particularly at low-$Z$ for aluminium and magnesium.
These predicted abundances are broadly consistent with the measurements
for EHe and RCB stars in the case of sodium and aluminium.  
The under abundance of magnesium in both EHe and RCB stars at low-$Z$ requires further investigation. 

\noindent {\it Phosphorus:} a primary motivation for this
investigation, the recipe shows that {\it \iso{31}{P} generated in an
AGB intershell can propagate to the surface of a subsequent 
white dwarf merger.} The recipe only predicts significant
overabundances at low-$Z$. Although this is consistent with 
one low-$Z$ phosphorus measurement and eight high-$Z$ EHe measurements, 
the recipe does {it not} explain a $\geq 1$ dex overabundances observed
in eight other high-$Z$ and one low-$Z$ measurements. 

\noindent {\it  Silicon, Sulphur, Argon, Calcium, Titanium, Chromium, and Nickel:} the
          {\it cold} merger recipe predicts negligible (silicon) or
          {\it no enhancement} beyond that expected from the
          enhancement of $\alpha$ elements in  low-$Z$ progenitors. 
The observations are {\it broadly} consistent with the recipe predictions for
all of these elements. A number of stars with up to 1 dex
enhancements of silicon, sulphur and/or argon demand further
attention. In particular, the observed scatter of $\pm1$ dex in
silicon is not obviously explained by this recipe. 

\noindent {\it s-process:} our provisional predictions for the
s-process elements  zinc, yttrium, zirconium, barium,
lanthanum, cerium and neodymium indicate a substantial excess is
expected in all cases except zinc. The predictions 
are all in excess of the mean trend of the observed abundances (or
their upper limits). For Y, Zr, and La, they are  
not in excess of the upper limit of the observed abundances. 
Given the small number of models available, and the probability that 
other factors will affect the observed distribution of s-process 
abundances, these results are inconclusive, but encouraging. 

It will be noted ({\it e.g.} from Fig.~\ref{f:modvobs}) that the
predicted excesses of certain elements are strong functions of the 
progenitor mass ($m_{\rm agb}$) and metallicity. 
Notable amongst these are neon, fluorine, sodium, and magnesium. 
An early objective of this investigation was
to determine whether the observed abundances placed any firm
constraints on the progenitor mass. For example, significant 
excesses (which are observed) in  neon, fluorine, and sodium 
are predicted for $1.9 < m_{\rm agb}/\Msolar < 3$, whilst an
excess of magnesium (which is not observed) is only predicted for 
$m_{\rm agb}/\Msolar \geq 3$. The evidence from phosphorus and
aluminium remains ambiguous. The suggestion is therefore that
the progenitor AGB stars had initial masses in the range 
$1.9 < m_{\rm agb}/\Msolar < 3$. This suggestion assumes that the 
binary components evolved essentially as single stars up to the AGB, 
plausible if the first common-envelope phase required to produce 
the short-period double-white-dwarf binary occurred after the more
massive star reached the AGB. 

\subsection{Cold merger: core mixing}
\label{s:cold:mixing}

Figure~\ref{f:alphamix} illustrates the effect of $\alpha_{\rm mix}$
on the abundances of the principal elements C, N, O and Ne, and also
shows the contribution of \iso{18}{O} to the total oxygen
abundance. It must be noted that these results depend strongly on the
composition profile at the carbon-helium boundary, and in particular
on the \iso{16}{O}:\iso{12}{C} ratio immediately below the boundary
layer. The intent is to demonstrate the effect of increasing the 
depth of mixing on the final model abundances.  

\noindent $\alpha_{\rm mix}=0$: the first row of Fig.~\ref{f:alphamix}
represents the case of no contribution from the boundary layer, oxygen
completely fails to match the observed abundances of EHe and RCrB
stars. There may be sufficient carbon in the intershell of
intermediate mass ($m_{\rm agb} \approx 2-3\Msolar$) to reproduce the
lower envelope of the carbon and neon abundances, but not the full
range.

\noindent $\alpha_{\rm mix}=1$: the second row of
Fig.~\ref{f:alphamix}, representing mixing down to the helium/carbon
equilibrium point, has similar results for carbon, nitrogen and neon,
but immediately shows much more oxygen at high $Z$. This oxygen is
almost entirely \iso{18}{O} (assuming that it survives the
actual merger).  The careful reader will note that the carbon
abundances are slightly depressed compared with $\alpha_{\rm mix}=0$; 
this is due to a mismatch between the intershell carbon abundances given in the
\citet{karakas10} models and the abundances in the pre-WD model
shown in Fig.\,4.

\noindent $\alpha_{\rm mix}=2$: slightly deeper mixing substantially
improves the carbon result without much change to other elements.

\noindent $\alpha_{\rm mix}>>2$: with very deep mixing, 
the models start to show too much carbon. Oxygen 
is increasingly composed of \iso{16}{O} and becomes independent of 
$Z$, although at high $Z$, \iso{18}{O} remains a significant
constituent. The contribution of \iso{22}{Ne} from the core has 
an almost negligible impact, even in high-$Z$ models. 

As noted above, a shift in the \iso{12}{C}/\iso{16}{O} ratio immediately below
the boundary layer alters the chemical balance. By setting 
$f_{\rm CO}=0.3$, it was possible to obtain a good correspondence
between model and observation for C, N, O, and the \iso{18}{O}/\iso{16}{O}
ratio at the same time (not shown). However, the value of $f_{\rm CO}$ required is 
unjustifiably smaller than the value of 0.5 obtained in the 
\citet{karakas10} models. The prospects for neon seem less
good; its abundance is primarily limited by the original CNO
abundance.

Further investigation of 
the chemical structure of white dwarf models derived from 
realistic AGB calculations will indicate whether the 
{\it cold} merger model stands up to scrutiny. Meanwhile,
Fig.~\ref{f:alphamix} suggests that $\alpha_{\rm mix}=3$ is 
satisfactory for the time being. Moreover, by indicating how much
material from below the C/He boundary must have been mixed, 
the value of  $\alpha_{\rm mix}$ may also tell us something 
about the ({\it cold}) merger process itself.  

\subsection{Hot merger} 

Figure~\ref{f:hotvobs} shows the chemical yields predicted by this
simplification of the {\it hot merger} model. In this case, there
is {\it no} contribution to the carbon and oxygen from mixing with the 
He/CO boundary layer from the COWD. {\it All} of the excess
carbon and oxygen comes from nucleosynthesis during the merger. 

The calculations by \citet{loren09} indicate significant production of
iron, nickel and zinc, and very large quantities of argon, calcium,
titanium and chromium produced by $\alpha$-capture reactions 
in the hot corona, especially towards the
upper limit of the He-WD mass range ($0.4 \Msolar$). At this limit, 
the titanium and chromium yields are so high as to be off scale 
in Figure~\ref{f:hotvobs}; thus we have restricted the nucleosynthesis 
of these two elements to the lower limit of their predicted range. 
The theoretical yields of iron, nickel and zinc are below the threshold defined by
the initial metallicity and hence have no effect on the final
abundances (Fig.~\ref{f:hotvobs}) compared with the model for 
the {\it cold} merger (Fig.~\ref{f:modvobs}). 

The surface abundances following a {\it hot merger} as predicted by
the {\it simple recipe} are shown in Fig.~\ref{f:hotvobs}. Recall that 
mixing at the carbon-helium boundary is switched off in this
case. The predictions are complicated by having very few models
amongst which to interpolate, making the
$Z$-distribution of elements strongly affected by merger
nucleosynthesis indicative rather realistic ({\it cf.} argon and calcium)

\noindent {\it Carbon and Oxygen:}, all surface carbon and oxygen 
is produced by nucleosynthesis during the hot merger. The quantities
are comparable with those from the {\it cold} merger, but in this case
{\it no} parameters were tuned to achieve the correct outcome.  

\noindent {\it Sulphur:} additional sulphur produced in the {\it hot merger} 
provides a better fit to the observations than in the {\it cold} merger. 

\noindent {\it Argon:} there is weak evidence of an argon excess in
some EHes; this provides some support for its formation in a hot
merger, possibly in higher-mass mergers ($m_{\rm agb}>2\Msolar$). 

\noindent {\it Calcium, Titanium, and Chromium: } the SPH merger
calculations of \citet{loren09}, combined with the merger recipe
described here, predict very high surface abundances of these
three elements in nearly all cases.  There is no observational
evidence that any of these elements is significantly overabundant in
any EHe or RCB star analysed to date. If EHes and RCBs are formed in a
merger, there is no evidence that reactions leading to the 
production of calcium, titanium, or chromium operate during the merger
process. This places strong constraints on temperatures and timescale
of the {\it hot merger}.

\noindent {\it Fluorine:} although not treated in the models of 
\citet{loren09}, any \iso{18}{O} produced or mixed into 
the heated material will be at least partially burnt to make
\iso{19}{F}, providing the temperature does not significantly exceed 
$3.10^8$\,K, where the F will subsequently be destroyed by $\alpha$
captures. Hence the presence of any \iso{18}{O} will almost inevitably 
result in an excess of \iso{19}{F}. 

\noindent {\it Lithium:} Li should be destroyed duuing a hot merger. 
 \citet{loren09} do not include Li, so predictions for Li are not 
 shown in  Fig.~\ref{f:hotvobs}.

\subsection{Uncertainties}
\label{s:errors} 

The obvious question is what confidence can be placed on
using a simple recipe to predict the outcome of a 
complicated process? For any element which is produced outside the
standard hydrogen and helium-burning reactions ({\it e.g.} s-process
  elements in the AGB precursor), the major impact on
the final surface yield is the degree of dilution by the accreted
helium white dwarf. In our recipe this is primarily constrained by the
minimum white dwarf mass for a conservative merger. Were $m_{\rm He}$ 
to exceed this value, the predicted excesses would be reduced. This does
not help to explain neon, fluorine or phosphorus in EHes. Were the merger to
be non-conservative (efficiency $\alpha < 1$),  
yields might increase by factors $1/\alpha$. SPH calculations support
$\alpha = 1$. Efficiencies $\alpha \ll 1$ would be surprising. 
Explaining neon, fluorine and phosphorus would require $\alpha < 0.1$. 

The neglect of specific mixing processes or individual reactions in
the stellar evolution models will impact on the recipe predictions. 
For example, the intershsell compositions from \citet{karakas10} 
were computed without the addition of a $^{13}$C pocket. 
This pocket is thought to form by the mixing of protons from the 
H-rich envelope into the intershell during the deepest
extent of a third dredge-up episode. These protons are quickly captured by 
the abundant  $^{12}$C 
resulting in the formation of a $^{13}$C pocket. 
Here, neutrons are liberated by the reaction
 $^{13}{\rm C}(\alpha,n)^{16}{\rm O}$ 
during the interpulse. Note that in most calculations
the $^{13}$C pocket is added artificially or induced through the inclusion
of convective overshoot \citep[{\it e.g.}][]{herwig00}. 
Whilst $^{13}$C pockets facilitate the formation of s-process elements, 
they are also important for enhancing
the abundance of some lighter elements including
 $^{19}$F,  $^{22}$Ne, and  $^{23}$Na \citep{lugaro04,karakas10}. 

Legitimate questions concern the core and shell masses for the 
AGB stars, the yields obtained in the AGB intershell nucleosynthesis
and in the hot merger nucleosynthesis. For example, how sensitive are
the masses adopted here to the microphysics, {\it e.g.} 
convection, rotation, mass-loss and/or mass
transfer? Since the model abundances are computed for CO+He WDs with 
$q=q_{\rm crit}$, would additional dilution produced $q>q_{\rm crit}$
seriously compromise the results?
If mass transfer occurs before AGB evolution is complete,
or even before the star reaches the AGB, are the intershell-yield
versus initial-mass relations still acceptable? What would the
hot-merger nuclear-hydro calculations look like with more realistic
networks and initial composition? We have discussed the question of
the chemical structure of COWDs, particularly at the C/He boundary
layer. Additional calculations for {\it hot mergers} with more 
extensive reaction networks and robust starting mixtures are also urgently
needed.


\section{Conclusion}
\label{s:conc}

The object of this paper was to clarify the surface abundances which
might arise under conservative assumptions for the merger of a helium
white dwarf with a carbon-oxygen white dwarf, and to compare these
with observed abundance anomalies for extreme helium and R\,Coronae
Borealis stars.  We have collated the observational
data describing the surface abundances for the latter, and presented
it, element by element, in a way that demonstrates any current excess
over the progenitor composition. We have discussed the background to
the physics of binary white dwarf mergers and their association with
the formation of extreme helium stars and R\,Coronae Borealis stars. 
We have developed a more elaborate version of the {\it simple
  recipe} used to estimate surface abundances in previous discussions of
this question \citep{saio02,pandey06}. We have incorporated
state-of-the-art calculations of the masses and light-element composition of 
AGB intershell regions \citep{karakas10}. 
We have made allowance for the existence of an
\iso{18}{O} pocket at the outer edge of the CO core. We have
considered the difference between a {\it cold} and a {\it hot} merger,
{\it i.e.} whether additional nucleosynthesis occurs during the
destruction of the helium white dwarf. 

Both models successfully match, or can be made to match, the observed
surface abundances of carbon, nitrogen, and oxygen.
The excess nitrogen comes primarily from the helium white dwarf as the
residue of CNO-processed carbon and oxygen. 
In the case of  the {\it cold} merger model, the excess carbon and oxygen
comes from the carbon-helium boundary of the carbon-oxygen white
dwarf.  A substantial fraction of oxygen probably takes the form of
\iso{18}{O} from a pocket just beneath this boundary, 
but \iso{16}{O} dredged from deeper layers will also be present. 
In the case of the {\it hot} merger model, the excess carbon and oxygen can
be produced during the merger. It is possible that the observed excess
comes from both sources. 

Both models predict up to 1 dex enhancements of \iso{19}{F}
and \iso{31}{P}, but not enough to match the observed overabundances
of fluorine and phosphorus. These elements come from the 
AGB intershell. An examination of mixing at the carbon/helium boundary
of the COWD suggests that observed neon may come from the outer part
of the carbon/oxygen core.  

Both models predict modest overabundances of sodium, aluminium and magnesium, 
particularly at low metallicity. These broadly match the observed
abundance distributions in sodium and aluminium. Magnesium is not
observed in excess at low-$Z$. 

The {\it hot} merger model currently predicts overabundances of calcium,
titanium and chromium which are not observed. The model
overabundance of argon is roughly consistent with available 
measurements, suggesting that argon might be produced during hot
merger nucleosynthesis. We do not yet have AGB intershell
yields for argon, which would affect the {\it cold} merger
predictions. 

We still require state-of-the-art data for s-process yields in AGB
intershell. 

Overall, the majority of species observed to be overabundant in EHe
and RCB stars are found to be enhanced in reasonable quantity in 
one or other of the white dwarfs prior to merger. In other words,
additional nucleosynthesis during a merger solves very few problems,
although that does not mean that it does not happen.

\section*{Acknowledgments}
The authors gratefully acknowledge Detlef Sch\"onberner, Ulrich Heber, David Lambert, Gajendra Pandey, Kameswara
Rao, and Martin Asplund and the late Philip Hill for their inspirational efforts in
measuring stellar abundances over 50 years. 
They also thank John Lattanzio, Maria
Lugaro and Simon Campbell for discussions on AGB nuclear networks and
particularly the nuclear origin of phosphorus in EHe stars.

The Armagh Observatory is funded by direct grant from the Northern
Ireland Dept of Culture Arts and Leisure.

CSJ acknowledges an Australian National University visiting fellowship. 

\bibliographystyle{mn2e}
\bibliography{mnemonic,ehe}

\label{lastpage}
\end{document}